\documentclass[a4paper,fleqn,usenatbib]{mnras}

\usepackage{microtype}
\usepackage{graphicx}
\usepackage{booktabs}
\usepackage{newtxtext,newtxmath}
\usepackage[T1]{fontenc}

\newcommand{\Msun}{\,\mathrm{M}_\odot}
\newcommand{\MJ}{\,\mathrm{M}_\mathrm{J}}
\newcommand{\Rsun}{\,\mathrm{R}_\odot}
\newcommand{\RJ}{\,\mathrm{R}_\mathrm{J}}
\newcommand{\Lsun}{\,\mathrm{L}_\odot}
\newcommand{\uHz}{\,\mu\mathrm{Hz}}

\newcommand{\dd}{\,\mathrm{d}}
\newcommand{\Myr}{\,\mathrm{Myr}}
\newcommand{\Gyr}{\,\mathrm{Gyr}}
\newcommand{\K}{\,\mathrm{K}}
\newcommand{\AU}{\,\mathrm{AU}}
\newcommand{\kms}{\,\mathrm{km}\,\mathrm{s}^{-1}}
\newcommand{\eps}{\,\mathrm{e}^-\,\mathrm{s}^{-1}}
\newcommand{\degrees}{{^\circ}}
\newcommand{\mas}{\,\mathrm{mas}}
\newcommand{\dex}{\,\mathrm{dex}}
\newcommand{\FeH}{[\mathrm{Fe}/\mathrm{H}]}
\newcommand{\Teff}{T_\mathrm{eff}}

\newcommand{\st}[1]{_\mathrm{#1}}
\newcommand{\tn}[1]{$^\mathrm{#1}$} 

\newcommand{\thisstar}{HD~38529}
\newcommand{\corot}{CoRoT}
\newcommand{\Mfinal}{1.48\pm0.04\Msun}
\newcommand{\Rfinal}{2.68\pm0.03\Rsun}
\newcommand{\tfinal}{3.07\pm0.39\Gyr}

\title[Asteroseismic properties of HD~38529]
    {Robust asteroseismic properties of the bright
  planet host HD~38529}

\author[W.~H.~Ball et al.]{Warrick H. Ball,$^{1,2}$
  William J. Chaplin,$^{1,2}$
  Martin B. Nielsen,$^{1,2}$
  \newauthor
  Lucia Gonz{\'a}lez-Cuesta,$^{3,4}$
  Savita Mathur,$^{3,4}$
  {\^A}ngela R.~G.~Santos,$^{5}$
  Rafael Garc{\'i}a,$^{6,7}$
  \newauthor
  Derek Buzasi,$^{8}$
  Beno\^it Mosser,$^{9}$
  Morgan Deal,$^{10}$
  Amalie Stokholm,$^{2}$
  \newauthor
  Jakob R{\o}rsted Mosumgaard,$^{2}$
  Victor Silva Aguirre,$^{2}$
  Benard Nsamba,$^{11,10}$
  \newauthor
  Tiago Campante,$^{10,12}$
  Margarida S.~Cunha,$^{10,12,1}$
  Joel Ong,$^{13}$
  Sarbani Basu,$^{13}$
  \newauthor
  Sibel {\"O}rtel,$^{14}$
  Z.~{\c C}elik~Orhan,$^{14}$
  Mutlu Y{\i}ld{\i}z,$^{14}$
  Keivan Stassun,$^{15}$
  \newauthor
  Stephen R.~Kane$^{16}$,
  Daniel Huber$^{17}$
  \\
  $^{1}$School of Physics and Astronomy, University of Birmingham, Edgbaston, Birmingham B15 2TT, United Kingdom\\
  $^{2}$Stellar Astrophysics Centre, Department of Physics and Astronomy, Aarhus University, Ny Munkegade 120, DK-8000 Aarhus C, Denmark\\
  $^{3}$Instituto de Astrof{\'i}sica de Canarias, La Laguna, Tenerife, Spain\\
  $^{4}$Dpto. de Astrof{\'i}sica, Universidad de La Laguna, La Laguna, Tenerife, Spain\\
  $^{5}$Space Science Institute, 4765 Walnut Street, Suite B, Boulder CO 80301, USA\\
  $^{6}$IRFU, CEA, Universit{\'e} Paris-Saclay, F-91191 Gif-sur-Yvette, France\\
  $^{7}$AIM, CEA, CNRS, Universit{\'e} Paris-Saclay, Universit{\'e} Paris Diderot, Sorbonne Paris Cit{\'e}, F-91191 Gif-sur-Yvette, France\\
  $^{8}$Department of Chemistry and Physics, Florida Gulf Coast University, 10501 FGCU Blvd., Fort Myers, FL 33965 USA\\
  $^{9}$LESIA, Observatoire de Paris, Universit{\'e} PSL, CNRS, Sorbonne Universit{\'e}, Universit{\'e} de Paris, 92195 Meudon, France\\
  $^{10}$Instituto de Astrof{\'i}sica e Ci{\^e}ncias do Espa{\c c}o, Universidade do Porto, Rua das Estrelas, PT4150-762 Porto, Portugal\\
  $^{11}$Max-Planck-Institut f\"{u}r Astrophysik, Karl-Schwarzschild-Str. 1, D-85748 Garching, Germany\\
  $^{12}$Departamento de F{\'i}sica e Astronomia, Faculdade de Ci{\^e}ncias da Universidade do Porto, Rua do Campo Alegre, s/n, PT4169-007 Porto, Portugal\\
  $^{13}$Department of Astronomy, Yale University, P.O. Box 208101, New Haven, CT 06520-8101, USA\\
  $^{14}$Department of Astronomy and Space Sciences, Science Faculty, Ege University, 35100, Bornova, {\.I}zmir, Turkey\\
  $^{15}$Department of Physics \& Astronomy, Vanderbilt University, Nashville, TN 37235, USA\\
  $^{16}$Department of Earth and Planetary Sciences, University of California Riverside, 900 University Ave, Riverside, CA 92521, USA\\
  $^{17}$Institute for Astronomy, University of Hawai‘i, 2680 Woodlawn Drive, Honolulu, HI 96822, USA
}

\date{Accepted 2020 October 9. Received 2020 September 11; in original form 2020 March 27.}
\pubyear{2020}

\begin{document}
\label{firstpage}
\pagerange{\pageref{firstpage}--\pageref{lastpage}}
\maketitle

\begin{abstract}
  The \emph{Transiting Exoplanet Survey Satellite} (TESS) is recording
  short-cadence, high duty-cycle timeseries across most of the sky, which
  presents the opportunity to detect and study oscillations in
  interesting stars, in particular planet hosts.  We have detected and
  analysed solar-like oscillations in the bright G4 subgiant
  \thisstar{}, which hosts an inner, roughly Jupiter-mass planet on a
  $14.3\dd$ orbit and an outer, low-mass brown dwarf on a $2136\dd$
  orbit.  We combine results from multiple stellar modelling teams to
  produce robust asteroseismic estimates of the star's properties,
  including its mass $M=\Mfinal{}$, radius $R=\Rfinal{}$ and age
  $t=\tfinal{}$.  Our results confirm that \thisstar{} has a mass near
  the higher end of the range that can be found in the literature and
  also demonstrate that precise stellar properties can be measured
  given shorter timeseries than produced by \corot{}, \emph{Kepler} or
  \emph{K2}.
\end{abstract}


\section{Introduction}

Stellar oscillations are sensitive to many of a star's basic
mechanical properties (e.g. its mass $M$ and radius $R$)
and can be measured very precisely.  The study
of these oscillations---\emph{asteroseismology}---thus provides a
precise tool with which to infer these mechanical properties, which
are in turn related to other important properties like a star's age.
Recently, the field has benefitted from a series of space missions
that recorded precise photometric timeseries: \corot{} \citep{baglin2006,corot2016},
\emph{Kepler} \citep{kepler} and \emph{K2} \citep{k2}.  They have 
revolutionised the study of solar-like oscillations
\citep[see e.g.][]{hekker2017,garcia2019}, which are
stochastic oscillations in cool stars, excited and damped by
near-surface convection across a large frequency range.  The
intrinsically low amplitudes, short lifetimes and incoherent phases of
solar-like oscillations makes them difficult to study from the ground
but the nearly-uninterrupted, short-cadence space-based observations
by \corot{}, \emph{Kepler} and \emph{K2} avoided these issues.

These missions were restricted to selected targets in a
number of relatively small fields of view, so the benefits of the
modern era of asteroseismology have been limited to these fields too.
The \emph{Transiting Exoplanet Survey Satellite} (TESS) has
been recording photometric timeseries that cover most of the sky since
July 2018.  Though TESS's photometry is less precise
than \corot{}'s or \emph{Kepler}'s at a given magnitude, it
presents the opportunity to apply the methods of asteroseismology to
bright, otherwise interesting solar-like oscillators whose
oscillations have
not been studied before \citep[e.g.][]{campante2019,nielsen2020}.

\thisstar{} (HR~1988, TIC~200093173) is a bright ($G=5.7332$) G4 subgiant,
around which
\citet{fischer2001} discovered a close companion with minimum mass
$M\st{b}\sin i\approx0.8\MJ$ on a $14.3\dd$ orbit.  They also reported
evidence of a more massive companion with an orbit exceeding
$1500\dd$, which they subsequently confirmed \citep{fischer2003} with
a period of about $2140\dd$ and minimum mass $M\st{c}\sin
i\approx13\MJ$.  Most recently, \citet{luhn2019} reported $M\st{b}\sin
i=0.797\pm0.15\MJ$ and $M\st{c}\sin i=12.99\pm0.15\MJ$, based on a
stellar mass $M=1.41\Msun$ \citep{brewer2016}.
\citet{benedict2010} combined radial
velocities with astrometric measurements from the Fine Guidance Sensor
aboard the Hubble Space Telescope to constrain the orbital inclination
of the outer companion to $i=47.3\pm3.7\degrees$.  They used
  the stellar mass estimate $M=1.48\pm0.05\Msun$ from \citet{takeda2007b}
  to infer that
the outer companion has a mass $M\st{c}=17.7\pm1.1\MJ$ and is
more massive than the brown dwarf lower-limit of about
$13\MJ$ \citep{spiegel2011}.
The system was monitored extensively by the \emph{Transit
    Ephemeris Refinement and Monitoring Survey}
  \citep[TERMS,][]{kane2009} whose long-term photometry ruled out
  transits by the inner planet \citep{henry2013}.

\begin{figure}
  \centering
  \includegraphics[width=\columnwidth]{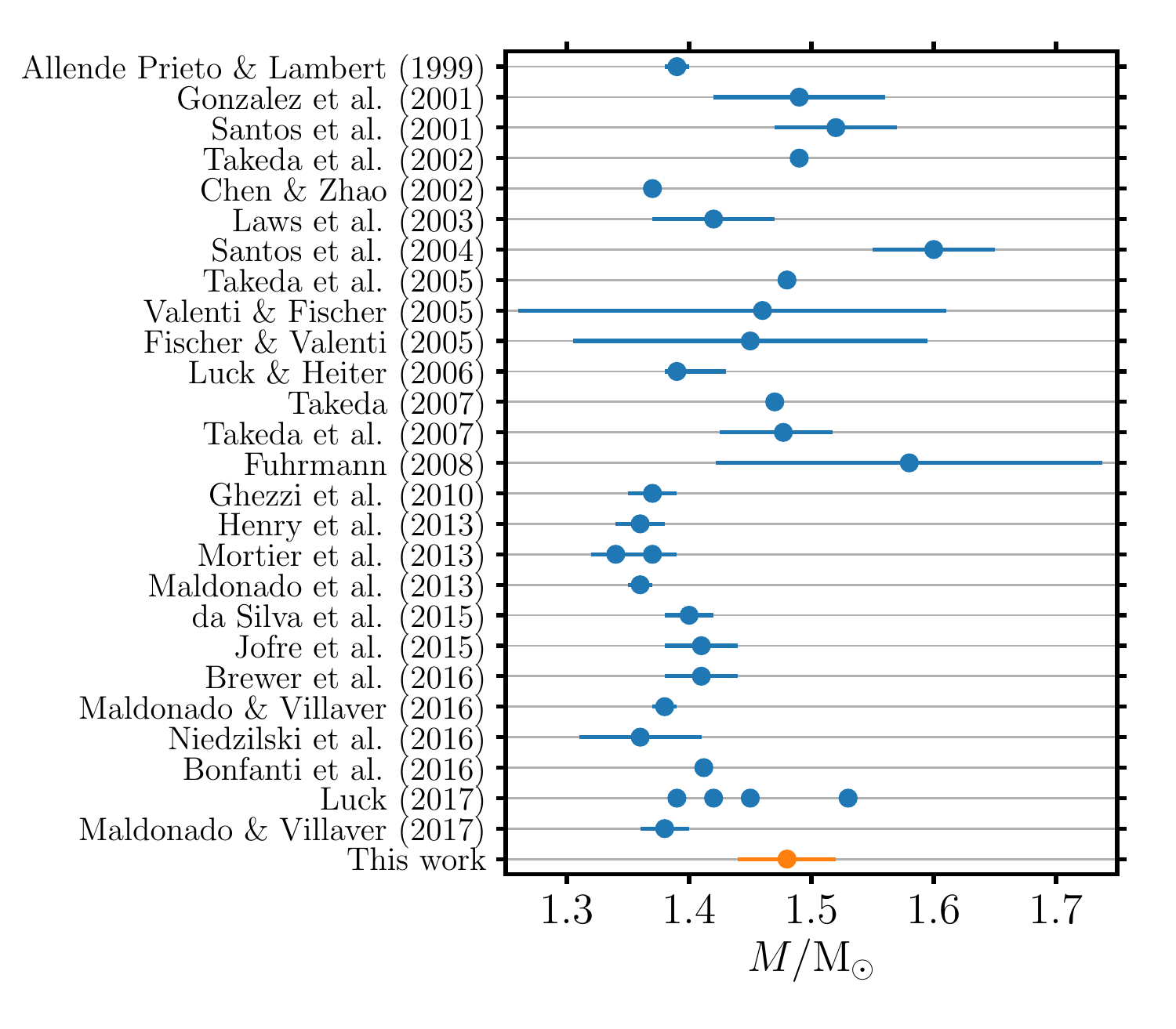}
  \caption{A selection of mass estimates for \thisstar{} from the
      literature, as well as the estimate from this paper.  Note that
      \citet{mortier2013} reported two different values estimated
      using different line lists for the spectroscopic parameters.
      \citet{luck2017} reported values for four different sets of
      isochrones.}
  \label{f:masses}
\end{figure}

As one of the just 59 planets discovered by the end of 2001 (according
to the NASA Exoplanet Archive), the system has been studied keenly
since and features in many exoplanet catalogues, surveys and archives.
Fig.~\ref{f:masses} shows a selection of masses from the literature,
many of which have been used in other articles.
Here, we fit a variety of stellar models to the observed spectrum of
solar-like oscillations to infer a robust asteroseismic mass for
\thisstar{} and also provide other asteroseismic properties, including
its radius and age.

\nocite{allendeprieto1999,
  gonzalez2001,
  santos2001,
  takeda2002,
  chen2002,
  laws2003,
  santos2004,
  takeda2005,
  valenti2005,
  fischer2005,
  luck2006,
  takeda2007a,
  takeda2007b,
  fuhrmann2008,
  ghezzi2010,
  henry2013,
  mortier2013,
  maldonado2013,
  dasilva2015,
  jofre2015,
  brewer2016,
  maldonado2016,
  niedzielski2016,
  bonfanti2016,
  luck2017,
  maldonado2017}

\begin{figure}
  \centering
  \includegraphics[width=\columnwidth]{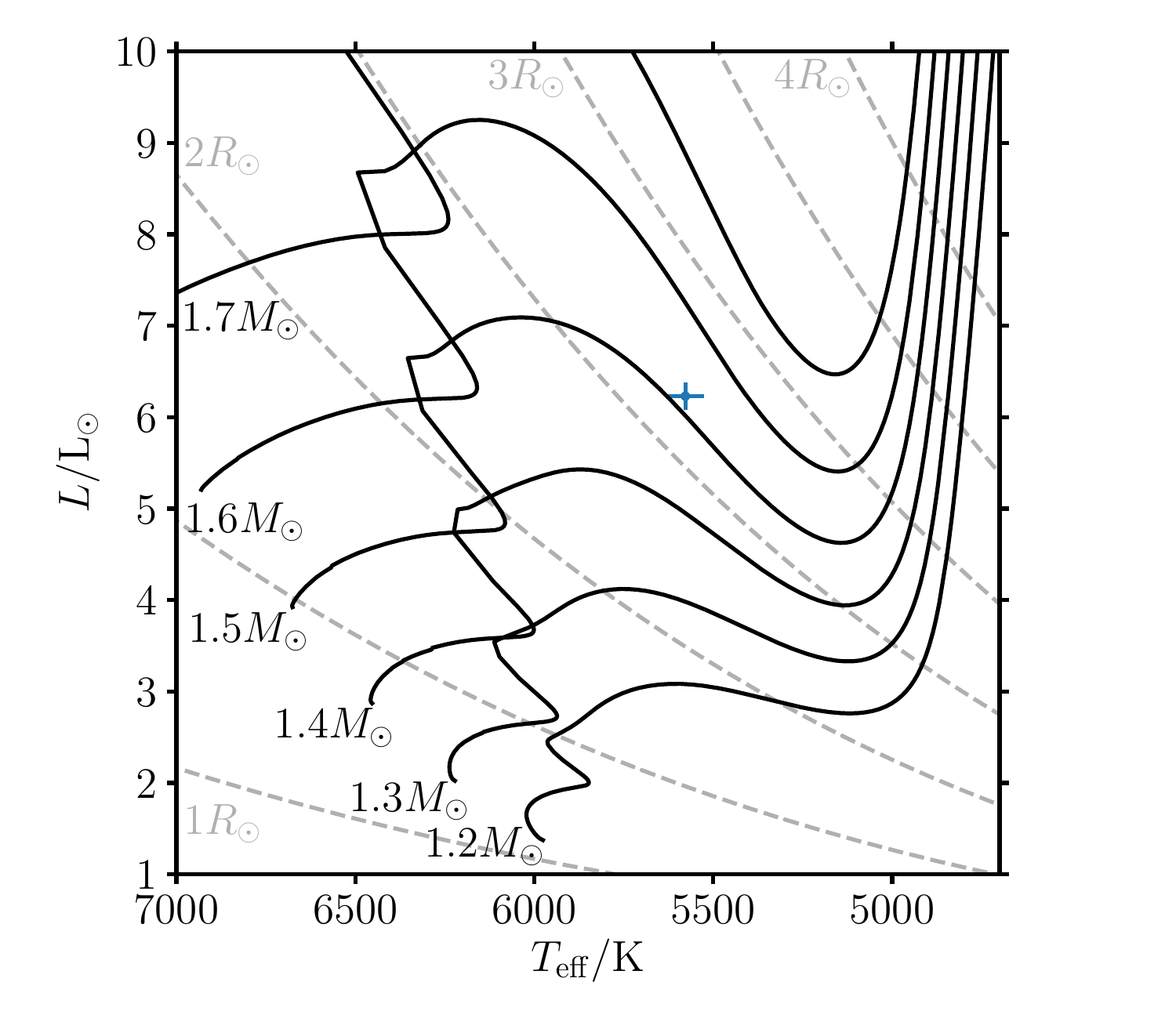}
  \caption{A Hertzsprung--Russell (HR) diagram showing the location of
    \thisstar{} (blue point).  The solid black lines show evolutionary
    tracks using the adopted
    metallicity $\FeH=0.34\dex$ and masses from $1.2$ to $1.7\Msun$ in
    steps of $0.1\Msun$.  The dashed grey lines are lines of constant
    radius from $1.0$ to $4.0\Rsun$ in steps of $0.5\Rsun$.}
  \label{f:HR}
\end{figure}

\begin{figure}
  \centering
  \includegraphics[width=\columnwidth]{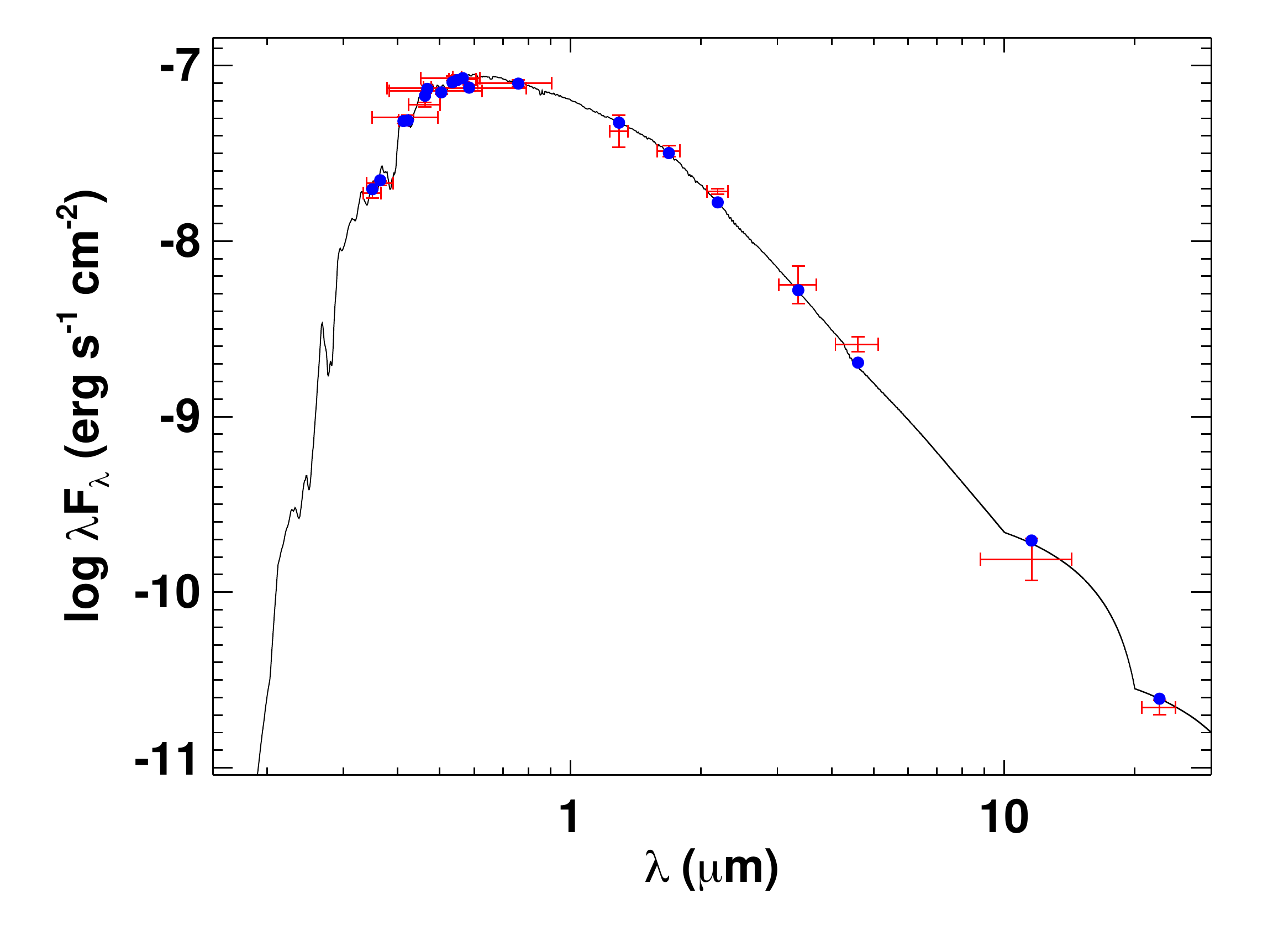}
  \caption{The spectral energy distribution (SED) of \thisstar{}.
    Data are indicated by red points and the best-fitting model by
    the solid black line.  The blue points are the model's integrated
    flux in the relevant filters.}
  \label{f:SED}
\end{figure}

\begin{table*}
  \centering
  \caption{Spectroscopic measurements.}
  \begin{tabular}{lcccc}
    \toprule
    Source & $\Teff/\K$ & $\FeH/\dex$ & $\log g/\dex$ \\
    \midrule
    \citet{deka2018}      & $5618 \pm 15$ & $0.38 \pm 0.02$ & $3.96 \pm 0.04$ \\
    \citet{maldonado2016} & $5585 \pm 18$ & $0.30 \pm 0.02$ & $3.86 \pm 0.05$ \\
    \citet{brewer2016}    & $5541 \pm 60$ & $0.32 \pm 0.06$ & $3.77 \pm 0.15$ \\
    \citet{jofre2015}     & $5573 \pm 31$ & $0.37 \pm 0.05$ & $3.81 \pm 0.03$ \\
    \citet{kang2011}      & $5574 \pm 74$ & $0.32 \pm 0.09$ & $3.76 \pm 0.10$ \\
    \midrule
    Adopted & $5578 \pm 52$ & $0.34 \pm 0.06$ & $3.83 \pm 0.11$ \\
    \bottomrule
  \end{tabular}
  \label{t:spec}
\end{table*}

\section{Observations}

\subsection{Non-seismic}
\label{ss:nonseismic}

We assembled a list of spectroscopic parameters determined using
different instruments and telescopes over the last ten years,
summarised in Table~\ref{t:spec}.  To combine these measurements into
a set of representative values, we averaged the means and
uncertainties and increased the uncertainties by the standard
deviation of the means, in quadrature.  This led to the adopted values
of $\Teff=5578\pm52\K$, $\FeH=0.34\pm0.06\dex$ and $\log g=3.83\pm0.11\dex$,
though the asteroseismic observations constrain $\log g$ much more
tightly than the spectroscopic value.  The measurements from the
individual sources are remarkably consistent, so the source of the
parameters is not decisive in our stellar model fits.
Fig.~\ref{f:HR} shows the location of \thisstar{} in a
Hertzsprung--Russell (HR) diagram, using the luminosity derived
in the next paragraph.  \thisstar{} is clearly
a slightly evolved, metal-rich subgiant.

We derived a bolometric luminosity by fitting the spectral energy
distribution (SED) using the methods described by \citet{stassun2016}
and \citet{stassun2017,stassun2018}.  Photometry is available for
photometric bands that cover
wavelengths from $0.35$ to $22\mu\mathrm{m}$, as shown in
Fig.~\ref{f:SED}.  The specific sources are
homogenised $UBV$ magnitudes from \citep{mermilliod2006},
$B\st{T}V\st{T}$ magnitudes from Tycho-2 \citep{tycho2a,tycho2b},
Str{\"o}mgren $uvby$ magnitudes from \citep{paunzen2015},
$JHK\st{S}$ magnitudes from 2MASS,
$W1$--$4$ magnitudes from WISE \citep{wise},
and Gaia's $G$, $G_{BP}$ and $G_{RP}$ magnitudes.
We fit the SED using the stellar atmosphere models
by \citet{kurucz2013} with priors on the effective temperature
$\Teff$, surface gravity $\log g$ and metallicity $\FeH$ from the
spectroscopic values above.  The extinction was fixed at zero because
of the star's small distance of $42.4\pm0.1\,\mathrm{pc}$
implied by its Gaia DR2 parallax of $23.582\pm0.059\mas$.
Integrating the model SED gives a bolometric
flux at the Earth $\mathcal{F}\st{bol}=(1.113\pm
0.026)\times10^{-7}\,\mathrm{erg}\,\mathrm{s}^{-1}\,\mathrm{cm}^{-2}$,
which, combined with the Gaia DR2 parallax, gives a bolometric
luminosity $L=6.23\pm0.15\Lsun$.  The best-fitting model is also shown in
Fig.~\ref{f:SED}.

\citet{baines2008a} and \citet{henry2013} both measured \thisstar{}'s
angular size using CHARA, finding mutually-consistent limb-darkened
angular sizes $\theta_\mathrm{LD}$ of $0.573\pm0.049\mas$ and
$0.593\pm0.016\mas$, respectively.  Given the Gaia DR2
parallax, these imply stellar radii of
$2.61\pm0.22\Rsun$ and $2.70\pm0.07\Rsun$.
Gaia DR2 includes a radius estimate of $2.81_{-0.21}^{+0.09}\Rsun$,
based on the $G$, $G_{BP}$ and $G_{RP}$ magnitudes \citep{andrae2018}.
The radius is degenerate
with $L$ and $\Teff$ when fitting our stellar models so we did not use
it as a constraint, though we do compare our best-fitting radius with
these independent values.

\begin{figure}
  \centering
  \includegraphics[width=\columnwidth]{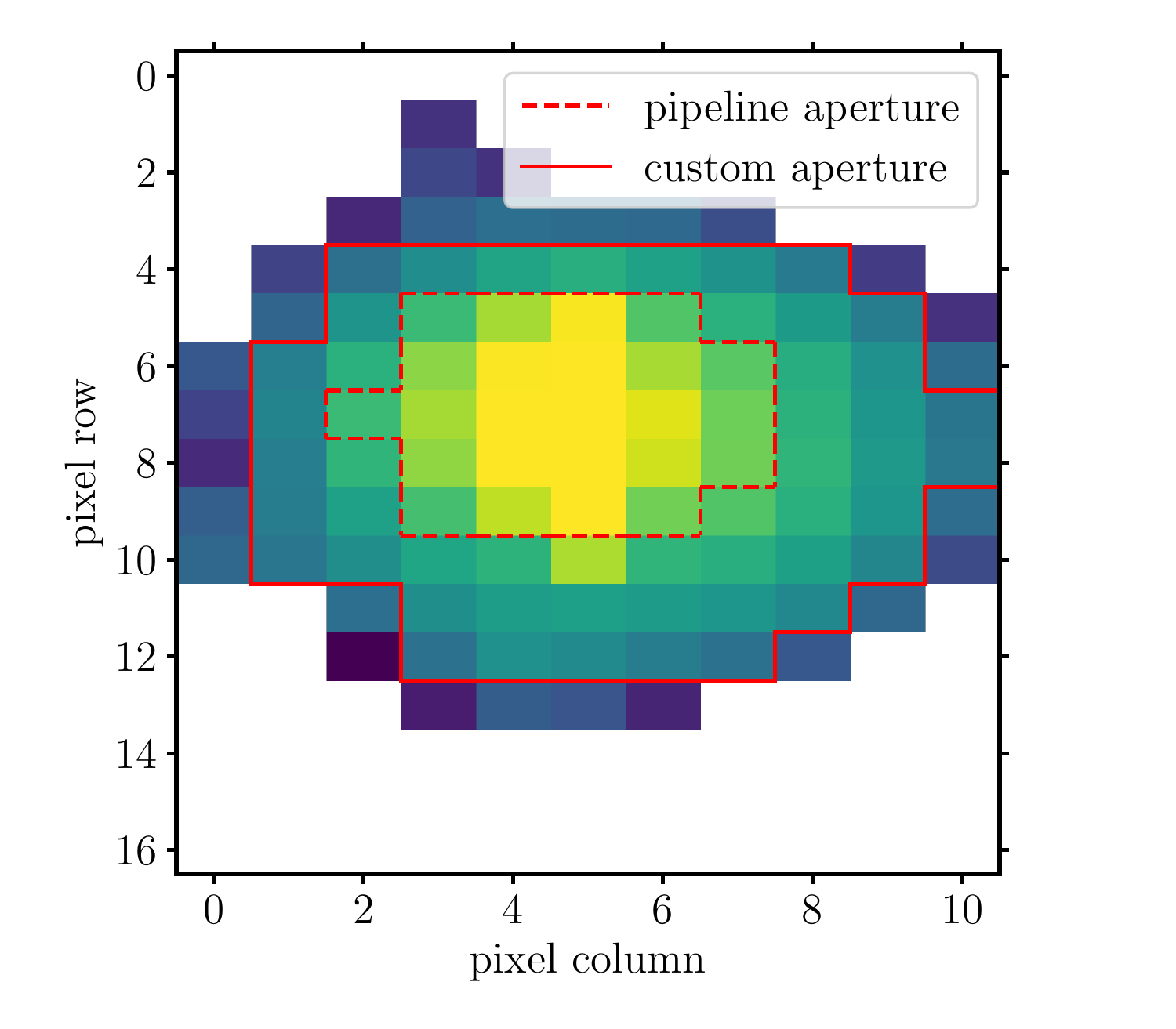}
  \caption{Median image of \thisstar{} during TESS's Sector 6
    observations, with a logarithmic colour scale.  The dashed and
    solid red lines show the default pipeline aperture and our custom
    aperture, respectively.  White regions had negative median
    fluxes, which are possible because of the SPOC pipeline's
    background subtraction.}
  \label{f:aperture}
\end{figure}

\begin{figure}
  \centering
  \includegraphics[width=\columnwidth]{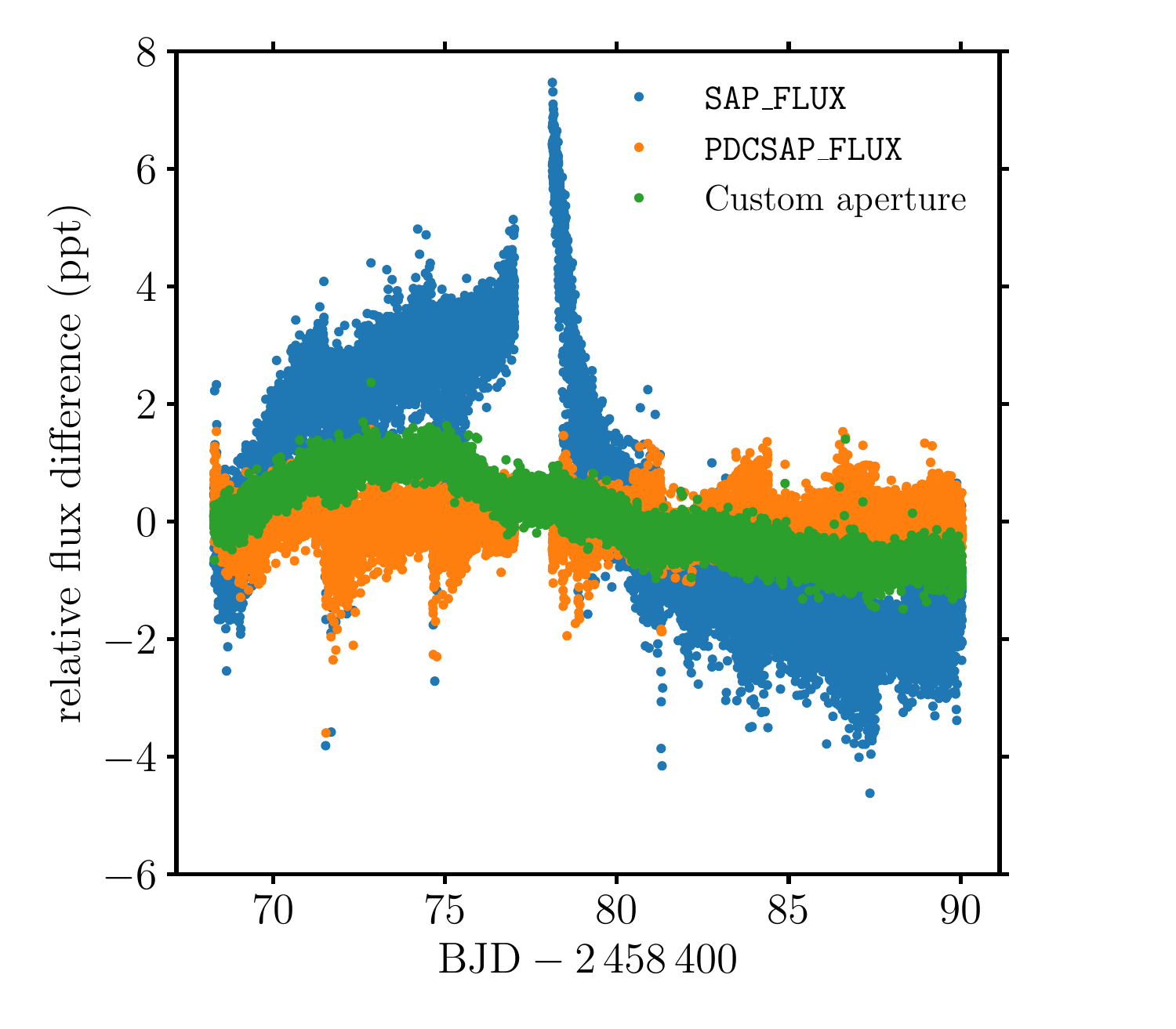}
  \caption{Lightcurves of \thisstar{} using either the default
    pipeline's \texttt{SAP\_FLUX} or \texttt{PDCSAP\_FLUX} data (blue
    or orange), compared with our custom, inpainted lightcurve (green).}
  \label{f:LC}
\end{figure}

\begin{figure}
  \centering
  \includegraphics[width=\columnwidth]{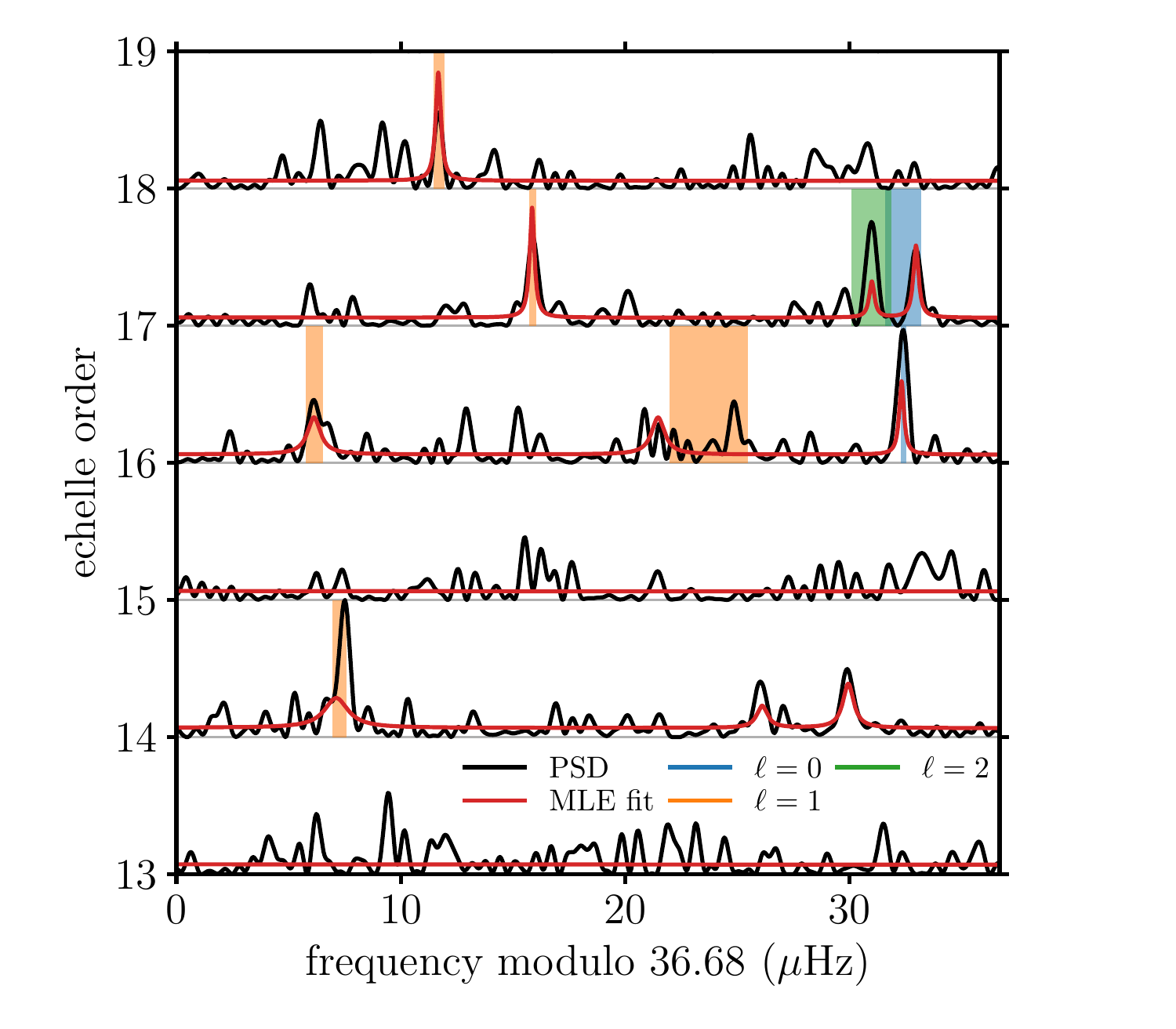}
  \caption{{\'E}chelle-like diagram of \thisstar{} computed using
      the Lomb--Scargle periodogram on the custom lightcurve in
      Fig.~\ref{f:LC} with an oversampling factor of 10.  The
      periodogram is shown as a black curve, normalized to its maximum
      value between $400$ and $700\uHz$.  The
      blue, orange and green areas show the $1\sigma$ uncertainty
      ranges covered by the $\ell=0$, $1$ and $2$ mode frequencies
      given to the stellar modelling teams (see Table~\ref{t:seis}).
      The red curve is the MLE fit by the Birmingham team.}
  \label{f:PS_stack}
\end{figure}

\begin{table}
  \centering
  \caption{Measured mode frequencies, all in $\uHz$.}
  \begin{tabular}{ccccc}
    \toprule
           & Paris   & Fort Myers & Birmingham & Adopted \\
    $\ell$ & UP+Asy. & Sig.~test & MLE \\
    \midrule
$0$ & $      $ & $543.46$ & $543.44 \pm 0.23$ \\
$0$ & $583.47$ & $      $ & $583.61 \pm 0.49$ \\
$0$ & $619.29$ & $619.28$ & $619.24 \pm 0.12$ & $619.27 \pm 0.12$ \\
$0$ & $654.85$ & $656.48$ & $656.53 \pm 0.21$ & $655.95 \pm 0.81$ \\
$1$ & $      $ & $486.29$ & $482.88 \pm 0.63$ \\
$1$ & $520.81$ & $520.65$ & $520.89 \pm 0.29$ & $520.78 \pm 0.31$ \\
$1$ & $592.98$ & $592.88$ & $593.22 \pm 0.36$ & $593.03 \pm 0.39$ \\
$1$ & $611.72$ & $611.70$ & $608.37 \pm 0.75$ & $610.59 \pm 1.74$ \\
$1$ & $639.43$ & $639.45$ & $639.44 \pm 0.16$ & $639.44 \pm 0.16$ \\
$1$ & $671.99$ & $671.91$ & $671.90 \pm 0.24$ & $671.93 \pm 0.24$ \\
$2$ & $539.55$ & $      $ & $539.55 \pm 0.25$ \\
$2$ & $654.51$ & $654.56$ & $654.49 \pm 0.89$ & $654.52 \pm 0.89$ \\
$2$ & $      $ & $669.43$ & $669.41 \pm 0.29$ \\    
    \bottomrule
  \end{tabular}
  \label{t:seis}
\end{table}

\subsection{Seismic}
\label{ss:seismic}

\thisstar{} was observed by TESS on its camera 1 during Sector 6 of
Cycle 1 (2018 December 15 to 2019 January 6).  We found no
oscillations in the SPOC pipeline lightcurves \citep{spoc}
despite the star being
among the top-ranked targets for asteroseismic detection in TESS's
\emph{Asteroseismic Target List} \citep[ATL,][]{schofield2019}.  We
therefore computed a custom lightcurve in which we expanded the
photometric aperture to include all pixels with a median flux greater
than $10$ electrons per second ($\eps$).  We found oscillations around
roughly $600\uHz$ in this custom lightcurve, though we note that the
ATL predicted the oscillations would peak around $400\uHz$.

To create
a suitable lightcurve for subsequent analysis,
we computed the total flux in apertures of different sizes.
We considered flux thresholds
starting from $10\eps$ for the largest aperture and increasing
progressively in increments of $10\eps$, $50\eps$, $100\eps$ and
$200\eps$, until reaching the standard TESS aperture, which is the
smallest aperture studied (see Gonz{\'a}lez-Cuesta et al., in prep.). For
all the apertures, we extracted the lightcurves and computed the power
spectrum density (PSD). Our seismically-optimised aperture is the one
where the oscillation modes' signal-to-noise ratio is highest.
We calibrated the lightcurve from the
optimised aperture using the Kepler Asteroseismic
Data Analysis Calibration Software \citep[KADACS,][]{garcia2011}
that was developed and tested on \emph{Kepler} data to remove outliers
and correct jumps. Finally, we filled the gaps with the inpainting
techniques by \citet{garcia2014} and \citet{pires2015}. For \thisstar{},
the optimal aperture was obtained with a flux threshold of $200\eps$.

The different apertures are shown in Fig.~\ref{f:aperture} and the
lightcurves in Fig.~\ref{f:LC}.  Both the standard pipeline
lightcurves (\texttt{SAP\_FLUX} and \texttt{PDCSAP\_FLUX}) have
increased scatter around the times of spacecraft thruster firings.
With our larger aperture, more of the star's light falls within the
aperture during these motions, rather than being lost as bright parts
of the star's point spread function move in and out of the aperture.

Fig.~\ref{f:PS_stack} shows an {\'e}chelle-like diagram of the power spectrum
in the region that includes the detected oscillation modes, along with
the individual mode frequencies that were used to model the star.
The individual mode
frequencies were measured from the power spectrum by three separate
teams, which we identify by their affiliations,  each using a
different method.  The first team (Paris) fit the universal
pattern by \citet{mosser2011up} to identify the radial and quadrupole
($\ell=0$ and $2$) modes and the asymptotic expression by
\citet{mosser2015} to identify dipole ($\ell=1$) modes, selecting the
nearest significant peaks in the power spectrum as the observed mode
frequencies.  The second team (Fort Myers) selected the peaks above a
signal-to-noise ratio of 4 and the third team (Birmingham) used maximum-likelihood
estimation (MLE) to fit Lorentzians to significant peaks in the power
spectrum.
The MLE fit is also shown in Fig.~\ref{f:PS_stack}.
Only the MLE fit returned straightforward uncertainties,
which are derived from the inverse of the Hessian matrix of the fit.

To combine the various results, we conservatively selected mode
frequencies only where all three teams reported a mode.  Our adopted
mean mode frequencies are the averages of the three teams' frequency values.
The adopted variances are the sum of the variances from the MLE fit
and the variance of the three means. i.e., the adopted uncertainty
is the sum, in quadrature, of the MLE uncertainty and the standard
deviation of the three teams' values.
Table~\ref{t:seis} lists all the
mode frequencies identified by at least two teams as well as the
adopted values that were provided to the stellar modellers.

The dipole modes are clearly mixed, i.e. the normally acoustic modes
have coupled to gravity modes deep in the star's interior, causing
them to deviate from the nearly-regular spacing that is expected of
purely acoustic modes.  In particular, there are two dipole modes in
{\'e}chelle order $16$, which is only possible if the modes are mixed.
Because mixed modes are partially sensitive to the properties of the
stellar core, they have distinct diagnostic properties compared
with purely acoustic modes.

\begin{table*}
\centering
\caption{Stellar model settings for the different teams.  The
    mixing-length parameter for the Birmingham models is a
    correction factor for the slight difference between the
    mixing-length formulations used in the stellar models and those
    used in the calibration by \citet{mosumgaard2018}.}
\begin{tabular}{cccc}
\hline
Team                    & Aarhus        & Birmingham             & Porto         \\
\hline                                                                           
Models                  & GARSTEC\tn{a} & MESA\tn{b} (r10398)    & MESA (r9793) \\
Oscillations            & ADIPLS\tn{c}  & GYRE\tn{d}             & GYRE          \\
High-$T$ opacities      & OPAL\tn{e}    & OPAL                   & OPAL          \\
Low-$T$ opacities       & F05\tn{f}     & F05                    & F05           \\
EoS                     & OPAL\tn{g}    & MESA/OPAL              & MESA/OPAL     \\
Solar mixture           & AGSS09\tn{h}  & GN93\tn{i}             & GS98\tn{j}    \\
Helium law ($Y=\ldots$) & $0.25$--$0.34$ & $1.289Z+0.248$        & $2Z+0.248$    \\
Nuclear reactions       & NACRE\tn{k+l,m} 
                        & NACRE\tn{k}
                        & NACRE\tn{k+n,o} \\ 
Atmosphere              & Eddington     & \citet{mosumgaard2018} & Eddington     \\
$\alpha_\mathrm{MLT}$   & $1.5$--$2.1$  & $1.037$*               & $1.3$--$2.9$ \\
Surface correction      & BG14-1\tn{p}  & BG14-1                 & \citet{sonoi2015} \\
Overshooting            & None          & Free                   & None          \\
\\
\hline
Team                    & Yale-M        & Yale-Y               \\
\hline                                                         
Models                  & MESA (r12115) & YREC                 \\
Oscillations            & GYRE          & \citet{antia1994}    \\
High-$T$ opacities      & OPAL          & OPAL                 \\
Low-$T$ opacities       & F05           & F05                  \\
EoS                     & MESA/OPAL     & OPAL                 \\
Solar mixture           & GS98          & GS98                 \\
Helium law ($Y=\ldots$) & $0.25$--$0.32$ & $0.248$--$0.328$    \\
Nuclear reactions       & NACRE\tn{k}   & Solar fusion I\tn{q} \\
Atmosphere              & Eddington     & Eddington            \\
$\alpha_\mathrm{MLT}$   & $1.83$        & $1.6$--$2.2$         \\
Surface correction      & BG14-2        & BG14-2               \\
Overshooting            & None          & $U(0,0.4)$           \\
\hline
\\
\end{tabular}
\\
\begin{tabular}{lll}
  \tn{a}\citet{weiss2008} &
  \tn{b}\citet{paxton2011,paxton2013,paxton2015} &
  \tn{c}\citet{adipls} \\
  \tn{d}\citet{gyre1,gyre2} &
  \tn{e}\citet{iglesias1993,iglesias1996} &
  \tn{f}\citet{ferguson2005} \\
  \tn{g}\citet{rogers2002} &
  \tn{h}\citet{asplund2009} &
  \tn{i}\citet{grevesse1993} \\
  \tn{j}\citet{grevesse1998} &
  \tn{k}\citet{angulo1999} &
  \tn{l}\citet{formicola2004} \\
  \tn{m}\citet{hammer2005} &
  \tn{n}\citet{imbriani2005} &
  \tn{o}\citet{kunz2002} \\
  \tn{p}\citet{ball2014} &
  \tn{q}\citet{adelberger1998} \\
\end{tabular}
\label{t:physics}
\end{table*}

\section{Stellar modelling}

Five teams, identified by their affiliations, analysed \thisstar{}
using a variety of stellar evolution and oscillation codes, with a
range of choices for various physical properties (sometimes referred
to as \emph{input physics}).  The main choices are shown in
Table~\ref{t:physics}.  In the rest of this section, we briefly
comment on some notable choices and describe the procedures that each
team used to find best-fitting model parameters and uncertainties.

The oscillation mode frequencies of calibrated solar models are known
to differ systematically from those of the Sun because of poor
modelling of the near-surface layers.  These differences, known as
\emph{surface effects} \citep[see][for a review]{ball2017kasc},
presumably affect all solar-like oscillators and must therefore be
corrected or removed to obtain unbiased model parameters.  All the
teams here have applied existing formulae to the uncorrected model
frequencies $\nu_{\mathrm{uncorr},i}$ to create the corrected model
frequencies $\nu_{\mathrm{corr},i}$ that are then compared with the
data.  Other methods can be used when no modes are mixed and more mode
frequencies are measured
\citep[e.g.][]{roxburgh2003,roxburgh2015,roxburgh2016}.

All teams combine the $\chi^2$ contributions of different observations,
for which it is useful to define the $\chi^2$ contribution of a particular quantity
$q$ by
\begin{equation}
  \chi^2_q=\left(\frac{q\st{obs}-q\st{mdl}}{\sigma_q}\right)^2\mathrm{,}
\end{equation}
where $q\st{obs}$, $q\st{mdl}$ and $\sigma_q$ are the observed value,
modelled value and observed uncertainty for the quantity $q$.  In addition,
many teams used the total $\chi^2$ of the oscillation mode frequencies,
\begin{equation}
  \chi^2\st{seis}=\sum_{i=1}^{N\st{seis}}
  \left(\frac{\nu\st{obs}-\nu_{\mathrm{corr},i}}{\sigma_{\nu_i}}\right)^2\mathrm{,}
\end{equation}
where $N\st{seis}=8$ is the total number of observed modes and
$\nu_{\mathrm{corr},i}$ is the $i$th surface-corrected model
frequency.  We also define the contribution
of the non-seismic observations by
\begin{equation}
  \chi^2\st{non-seis}=\chi^2_{\FeH} + \chi^2_{\Teff} + \chi^2_{L}\mathrm{.}
\end{equation}

We note that, as is common in one-dimensional stellar evolution codes,
none of the models included the potentially relevant effects of
rotation or radiative levitation, which we comment on further in
Sec.~\ref{ss:transport}.  Only the Yale-Y team used any gravitational
settling.

\subsection{Aarhus}

The Aarhus team used the Bayesian fitting code BASTA \citep{silva2015,
  legacy2} to sample stellar models on a pre-computed grid.  The grid
spanned masses from $1.30$ to $1.60\Msun$, mixing-length parameters
$\alpha\st{MLT}$ from $1.5$ to $2.1$, initial metallicities $\FeH$
from $0.2$ to $0.5\dex$ and initial helium abundances from $0.25$ to
$0.34$.  The parameters were sampled with 5000 evolutionary tracks
selected by Sobol quasi-random sampling.  BASTA uses Bayesian
inference to compute the marginalised posterior of any stellar
quantity by integrating over all models and applying weights
to handle non-uniform sampling in the volume of the parameter
space.
For example, more models are computed during rapid phases of
evolution.  Without weights, the results would be biased towards
these rapid phases, so a weight is applied to avoid this.
The value reported for each quantity is the median of the
posterior with the 16th and 84th percentiles.
The objective function is the
likelihood $\mathcal{L}\propto\exp(-\chi^2\st{tot}/2)$, with
\begin{equation}
  \chi^2\st{tot} = \frac{1}{N_{\mathrm{seis}} - 1}\chi^2\st{seis}
  + \chi^2\st{non-seis}\mathrm{.}
\end{equation}

As the star evolves and the modes become mixed, multiple non-radial
modes ($\ell > 0$) can occur between two consecutive radial ($\ell =
0$) modes.  To decide which modes in the model should be included in
the likelihood function, BASTA matches the modes in the models to the
observed modes based on their separation in frequency as well as the
mode inertias \citep{acdk2010}.

For a given angular degree $\ell>0$, suppose there are $n>1$ modelled
modes between two radial modes, and that these non-radial modes
have inertias $I_1<\ldots<I_n$.  We possibly do not observe
all the modes between the radial modes so have some number $m\leq n$
of observed modes, and must somehow choose which $m$ modelled
modes to compare to the observed modes.
The simplest method is to select the modelled modes with the lowest
inertias, as these are expected to have the highest amplitudes, but
small differences in inertia might lead to an incorrect selection.

Instead BASTA creates two inertia thresholds $a=I_m/10$ and $b=10\,I_m$,
where $I_m$ is the $m$th-smallest inertia of the modelled modes
between the two radial
modes.  It then subdivides the modelled modes into a set $A$ with inertias
less than $a$, set $B$ with inertias between $a$ and $b$, and set
$C$ with inertias greater than $b$.  These thresholds roughly
distinguish modes that are likely to be detected (set $A$),
those that are unlikely to be detected (set $C$) and those
somewhere between (set $B$).  The
values of $a$ and $b$ ensure that $A$ has fewer than $m$ elements and
$A\cup B$ has at least $m$ elements.
These thresholds are determined from experience
and have led to robust results in all their applications so far.
By selecting all modes in $A$ and
a subset of $B$ such that $m$ modes are chosen in total, the modes can
be matched one-to-one to the observed modes.
If there are no modes in $A$, all the modes are selected from $B$.
To decide which modes to select from $B$,
BASTA uses the subset of $B$ with the smallest total
absolute frequency difference between the observed and modelled modes
(i.e. the $\mathrm{L}_1$ norm).

\subsection{Birmingham}

The Birmingham team used Modules for Experiments in Stellar
Astrophysics \citep[MESA, r10398;][]{paxton2011,paxton2013,paxton2015}
with the atmosphere models and calibrated mixing-length parameters from
\citet{trampedach2014a,trampedach2014b} as implemented in
\citet{mosumgaard2018}. The mixing-length parameter in Table
\ref{t:physics} is the solar-calibrated correction factor that
accommodates slight differences between MESA's input physics and
mixing-length model and that of the simulations by
\citet{trampedach2014a,trampedach2014b}.  All other teams used grey
Eddington atmosphere models.

The Birmingham team optimised the mass $M$, initial metallicity
$\FeH_i$, overshoot parameter $\alpha\st{ov}$ and age $t$
to minimise the unweighted total squared differences
between the model and both the seismic and non-seismic data, i.e.
\begin{equation}
  \chi^2\st{tot}=\chi^2\st{seis}+\chi^2\st{non-seis}\mathrm{.}
\end{equation}
The overshooting parameter $\alpha\st{ov}$ is the number of pressure
scale-heights that are chemically mixed beyond the formal convective
boundaries.
The team optimised the parameters using a combination of a
downhill simplex \citep[i.e. Nelder--Mead method,][]{nelder1965} and
samples
drawn randomly within error ellipses around the best-fitting
parameters when the simplex stagnated.  Uncertainties were estimated
by finding the parameters of minimum-volume ellipsoids that
simultaneously bound all samples with
$0.25<\chi^2\st{tot}-\mathrm{min}(\chi^2\st{tot})<25$ when their
distance to the optimum is scaled by $\sqrt{\chi^2\st{tot}}$, as
described by \citet{ball2017}.
  
\subsection{Porto}

The Porto team used the software package \emph{Asteroseimic Inference
  on a Massive Scale} \citep[AIMS,][]{rendle2019}, which interpolates
stellar properties in a precomputed grid and estimates parameters and
their uncertainties by Markov Chain Monte Carlo (MCMC) sampling of a
chosen posterior distribution.

The sampled posterior comprises uniform priors in appropriate ranges and a likelihood
function defined as $\mathcal{L}\propto\exp(-\chi^2\st{tot}/2)$, where 
\begin{equation}
  \chi^2\st{tot} = \frac{3}{N\st{seis}}\chi^2\st{seis}
    + \chi^2\st{non-seis}\mathrm{,}
\end{equation}
where the factor $3$ is used to balance the seismic constraints
with the three non-seismic constraints.

For \thisstar{}, the posterior distributions appear to be dominated by
a single stellar model in the underlying grid, with a limited
contribution from a few other models and interpolation around those
models.  To compute more reliable uncertainties, we use the points
at which the cumulative distribution functions are
equal to $0.0013$ and $0.9987$, and divide this range by three.
These points correspond to the $3\sigma$ limits of a normal distribution,
in the same way that the 16th and 84th percentiles correspond to
the $1\sigma$ limits.

\subsection{Yale-M}

The Yale-M team used the parallel differential evolution algorithm
by \citet{tasoulis2004} as implemented in the Python package \textsc{Yabox} \citep{yabox} to
find the optimal values of the mass, initial helium abundance and
initial metallicity.  The mass was allowed to vary between
$1.39$ and $1.61\Msun$, the initial helium abundance between
$0.25$ and $0.32$ and the initial metal-to-hydrogen ratio $Z/X$
between $0.038$ and $0.05$, which were chosen based on an initial
rough optimisation using only the radial mode frequencies and
non-seismic constraints.

The objective function is a total sum of squared
differences $\chi^2\st{tot}$, defined by
\begin{equation}
  \chi^2\st{tot}=\frac{1}{N\st{seis}}\chi^2\st{seis}+\chi^2\st{3}
  +\chi^2\st{non-seis}
\end{equation}
with
\begin{equation}
  \chi^2\st{3}=\frac{1}{3}\sum_{i=1}^3\chi^2_{\nu_{\mathrm{uncorr},i}}\mathrm{,}
\end{equation}
where $\nu_{\mathrm{uncorr},i}$ is the uncorrected model frequency.
The extra term $\chi^2\st{3}$ is the reduced $\chi^2$ of the three lowest
frequency modes, before correction, which acts as a prior that prefers
those models for which the three lowest uncorrected mode frequencies
are similar to the observed mode frequencies.

All the models generated by the differential evolution were retained,
which in effect created a non-uniform grid of models.  The density of
models in each region of parameter space was sampled using a kernel density
estimator (KDE), which defines a prior for how likely each model was
in the absence of any observations.  The total $\chi^2\st{tot}$ was then transformed
into a likelihood $\mathcal{L}\propto\exp(-\chi^2\st{tot}/2)$ from which the means
and standard deviations could be estimated from the moments of the
resulting formal posterior distribution.

\subsection{Yale-Y}

The Yale-Y team constructed a grid of models spanning masses from
$1.40$ to $1.60\Msun$ in steps of $0.01\Msun$, mixing length
parameters $\alpha\st{MLT}$ from $1.6$ to $2.2$ in steps of $0.075$,
initial helium abundances $Y_i$ from $0.248$ to $0.328$ in steps of
$0.01$ and initial metallicities $\FeH_i$ from $0.260$ to $0.390\dex$ in
steps of $0.015\dex$.

Each model had a random core overshoot parameter $\alpha\st{ov}$
selected uniformly between $0$ and $0.4$, with overshooting modelled
in the same way as the Birmingham team.
The models included gravitational settling,
with an efficiency multiplied by the factor
$\exp[-(1/2)(M/\Msun-1.25)^2/0.085^2]$ to prevent the heavy elements
from completely draining from the surface during the main-sequence
(see Sec.~\ref{ss:transport}).

The relative likelihood of each model was computed using
$\mathcal{L}\propto\exp(-\chi^2\st{tot}/2)$, with
\begin{equation}
  \chi^2\st{tot}=\frac{1}{N\st{seis}}\chi^2\st{seis}+\chi^2\st{non-seis}
  \mathrm{.}
\end{equation}
The reported values are the medians and 16th and 84th percentiles of
the likelihoods marginalised over all other parameters.

\begin{table*}
  \centering
  \caption{Best-fitting stellar model parameters.}
  \begin{tabular}{clllll}
    \toprule
    Team          & $M/\Msun$ & $R/\Rsun$ & $t/\Gyr$ & $L/\Lsun$ & $\bar\rho/(\mathrm{g}\,\mathrm{cm}^3)$ \\
    \midrule
      Aarhus & $1.480_{-0.031}^{+0.067}$ & $2.677_{-0.027}^{+0.037}$ & $3.17_{-0.16}^{+0.10}$ & $6.11_{-0.10}^{+0.20}$ & $0.1094_{-0.0003}^{+0.0003}$ \\
  Birmingham & $1.439 \pm 0.024$         & $2.653 \pm 0.017$         & $3.29 \pm 0.08$        & $6.00 \pm 0.09$        & $0.1085 \pm 0.0006$          \\
       Porto & $1.492\pm0.007$           & $2.686 \pm 0.006$         & $2.89 \pm 0.03$        & $6.32 \pm 0.11$        & $0.1085 \pm 0.0006$          \\
      Yale-M & $1.498 \pm 0.047$         & $2.691 \pm 0.029$         & $2.81 \pm 0.02$        & $6.18 \pm 0.08$        & $0.1083 \pm 0.0003$          \\
      Yale-Y & $1.489 \pm 0.030$         & $2.685 \pm 0.024$         & $3.20 \pm 0.74$        & $6.17 \pm 0.15$        & $0.1065 \pm 0.0012$          \\
    \midrule
     Adopted & $1.479 \pm 0.037$         & $2.678 \pm 0.026$         & $3.07 \pm 0.39$        & $6.16 \pm 0.15$        & $0.1083 \pm 0.0012$          \\
    \bottomrule
  \end{tabular}
  \label{t:results}
\end{table*}

\section{Results and Discussion}

The stellar parameter values inferred by each team are given in
Table~\ref{t:results}, along with consolidated parameter values.  The
consolidated values are computed by combining the results from each
team using the same method as for the spectroscopic data.
The main results are the mass $M=\Mfinal{}$, radius $R=\Rfinal{}$ and age
$t=\tfinal{}$.
The mass is near the upper end of the range of masses that
have appeared in the literature and similar to the value $1.48\pm0.05\Msun$
determined by \citet{takeda2007b} and used by \citet{benedict2010}.
The radius is measured more precisely
than in any previous study and our result is consistent with both
the Gaia DR2 value \citep{andrae2018} and the
interferometric measurements by \citet{baines2008a} and
\citet{henry2013} when combined with the Gaia DR2 parallax.

A sixth team independently calibrated a stellar model to the
spectroscopic data and radial frequencies only and found a consistent
mass $M=1.48\Msun$, radius $R=2.68\Rsun$ and age $t=2.70\Gyr$.  This
model also used MESA (r10000), ADIPLS, the solar mixture of
\citet{asplund2009}, the surface correction by \citet{kjeldsen2008}
and input physics otherwise similar to that of the Porto and Yale-M
teams.

\subsection{Precise age estimates}

Several of the pipeline's age estimates appear unreasonably precise.  As
a reference, we first note that the uncertainty on any single
evolutionary track is very small because of how quickly the mode
frequencies change with age
\citep[see ][for a detailed discussion]{deheuvels2011}.
In \thisstar{}, the dipole modes can
change at about $3\uHz/\Myr$ and the fastest changing mode takes
about $0.1\Myr$ to evolve by $1\sigma$.  The reported age
uncertainties are therefore dominated by the correlation of age with
other parameters, notably the mass.
A star's main-sequence lifetime is roughly proportional to $M^{-3}$,
so we roughly expect the fractional age uncertainty to be about $3$
times the fractional mass uncertainty, though this does not account
for correlations with other parameters.  The Birmingham team's
estimate is about half this value and the Yale-M team's estimate even
smaller, even though the other parameter uncertainties seem
reasonable. e.g. because the mean density $\bar\rho$ is
very tightly constrained, the
fractional uncertainty on mass is about $3$ times that of the radius.

Such precise ages for subgiants and low-luminosity red giants have
been encountered before
\citep[e.g.][]{deheuvels2011,ball2017,stokholm2019,li2020}
but in most
cases, the mass uncertainties are sufficiently precise that the age
uncertainties are still consistent.  We note, however, that
\citet{stokholm2019} inferred very precise ages for the bright
subgiant HR~7322 (KIC~10005473) and discuss the constraining
power of its mixed modes in detail.
\citet{li2020} also report age uncertainties that are more precise
than the na\"ive estimate for the stars KIC 6766513, KIC 7199397, KIC
10147635, KIC 11193681 and KIC 11771760.  There is no obvious
connection between these stars other than their best-fitting masses
all being greater than $1.3\Msun$.  We also note that, at least in the
Birmingham team's models, the dipole-mode frequencies are all
increasing while the star's radius is staying roughly constant,
as \citet{stokholm2019} also found for HR~7322.  Because the star's
mean density is therefore roughly constant, one would expect
purely acoustic mode frequencies to be roughly constant too.  That
the dipole-mode frequencies are increasing implies that they are
undergoing avoided crossings driven by changes to the star's
internal structure, which might reduce the correlation with
other parameters that should dominate the age uncertainty.

It is not clear how additional free parameters (e.g. the
initial helium abundance $Y_i$ or mixing length parameter
$\alpha\st{MLT}$) affect the age uncertainties.  It is possible
for certain combinations of parameters to be required for better
fits to the data, which could confine the age by having it
(anti)correlate with multiple parameters such that the simple
estimate here---which assumes no correlations---is an overestimate.
Even so, the more uncertain estimate by the Yale-Y team and the
extra uncertainty from the spread of means (which contributes about
$0.2\Gyr$) means that our overall result is less certain than the lower
bound suggested by the simple relationship between mass and age.

\subsection{Neglected transport mechanisms}
\label{ss:transport}

\thisstar{}'s mass places it in a region where stellar models
typically neglect several potentially important processes that can
transport chemical species in the star.  On the other hand,
\thisstar{} has evolved far enough that the inward movement of the
convective envelope's inner boundary will have already erased the
signal of some chemical peculiarities that may have existed while the
star was on the main sequence.  At this point in the star's evolution,
roughly the outer half by mass is convective.  Even so, the extra
chemical transport processes may have affected the structure of the
star in ways that still affect its observable appearance.

The first such process is rotation.  \thisstar{} would have been an
early- to mid-F-type star ($\Teff\approx6700\K$) on the main-sequence,
so may have rotated relatively quickly.  Measurements of the star's
current $v\sin i$ in the literature show a large spread, so we use the
estimate of the rotation period $P=31.65\pm0.17\dd$ by
\citet{benedict2010} based on photometry from the Hubble Space
Telescope's Fine Guidance Sensor.  We note that they report an
amplitude of $0.15$ per cent for the rotational modulation, in which
case the amplitude and period are consistent with the roughly
sinusoidal variation in our custom TESS lightcurve.

Though our understanding of angular momentum transport in evolved
stars has been shown to lack some important process
\citep{eggenberger2012,marques2013}, the star's surface gravity $\log
g\approx3.75\dex$ places it around the point at which the radial
rotation profiles appear to first depart from solid-body rotation
\citep[see e.g.][]{deheuvels2014,spada2016}.  The star's main-sequence
radius grew from about $1.4\Rsun$ at zero age to about $2.1\Rsun$ at
terminal age so, assuming solid body rotation, its rotation period
would have increased from about $8.5$ to $19.1\dd$.  Equivalently, the
rotational velocity decreased from about $8.4$ to $5.6\kms$.  It is
thus unlikely that \thisstar{} rotated quickly on the main sequence,
so the chemical transport by rotation was probably modest.

The second process we have neglected (or, for the Yale-Y team,
suppressed) is chemical diffusion, which describes the separate
processes of gravitational settling and radiative levitation
\citep{michaud2015}.  As is
common when modelling stars more massive than about $1.2$ to
$1.3\Msun$, we have neglected or suppressed gravitational settling
because current models predict that heavier elements are completely
drained from the stellar surface, which is clearly at odds with
observations.  It is usually assumed that some competing transport
process prevents this from happening but its precise nature is still
unknown \citep[see e.g.~Sec.~6.2 of][]{salaris2017}.

Radiative levitation is a related process that raises heavier elements
towards the stellar atmosphere because they are subject to a greater
radiative force against gravity than lighter elements.
\citet{deal2018} showed that this is an important process when
inferring the properties of main-sequence stars.  \citet{deal2020}
further showed that modest rotation (about $30\kms$) is insufficient
to prevent a discernible effect on the stellar properties.  Given that
\thisstar{} probably rotated more slowly, it may have experienced
significant heavy element enhancement at its surface on the main
sequence, even if much of the effect has since been erased by the
growing convective envelope.

To roughly quantify the effect of these neglected processes, we
first computed evolutionary tracks up to the observed $\Teff=5578\K$
with $M=1.48\Msun$, $\FeH=0.34$ and a rotation rate of $5\dd$ at age
$10\Myr$ as described in \citet{deal2020}.
Each track used one of the following combinations of the
extra chemical transport processes above: rotation, gravitational
settling and radiative levitation; gravitational settling and
radiative levitation; only gravitational settling; and no extra
chemical transport.  The tracks show that gravitational settling
leads to a longer main-sequence lifetime and a brighter subgiant
phase, which in turn suggests that we have overestimated the star's mass
and underestimated its age.  Radiative levitation appears
to have little effect on the main-sequence evolution and
any abundance anomalies are erased by the convection zone
on the subgiant branch.

We then varied the input mass of the tracks with rotation,
gravitational settling and radiative levitation to find a model that
reached the same values of $\log g$ and $\Teff$ as the $1.48\Msun$
track with no extra chemical transport.  The best-fitting model by
this approximate method has a mass of $1.395\Msun$ and is $31$ per
cent older than the $1.48\Msun$ model without extra chemical
transport.  From the constraint of fixed $\log g$, the radius is about
$3.0$ per cent smaller, which is roughly a $3.1\sigma$ difference.
The mass, radius and age therefore differ by about $2.4$, $3.1$ and
$2.5\sigma$, respectively, when using our reported fractional
uncertainties.  Though this analysis only varies the mass and age and
does not use any seismic constraints, it demonstrates the potential
importance of gravitational settling and rotation when determining the
properties of stars like \thisstar{}.

\subsection{Implications for companion brown dwarf}

As noted earlier, \thisstar{} hosts a planet and brown dwarf, and our
results present a number of implications for these companions.
\citet{luhn2019} provide the most recent measurements and used a host
mass of $1.41\Msun$ determined by \citet{brewer2016}.  The companion
masses scale with $M^{2/3}$ so our inferred mass implies that the
companions are $3.2$ per cent larger than \citet{luhn2019} report.

Our revised stellar properties affect the extent of the habitable zone
\citep[HZ, e.g.][]{kasting1993,kopparapu2013,kopparapu2014} around
\thisstar{}.  \citet{kane2016} defined ``conservative'' (based on
runaway and maximum greenhouse models) and ``optimistic'' (based on
empirical data from Venus and Mars) HZ boundaries, both of which are
sensitive to small changes in stellar properties and their associated
uncertainties \citep{kane2014}. Our radius of $\Rfinal{}$ and adopted
effective temperature of $5578\pm52\K$ (see Sec.~\ref{ss:nonseismic})
result in calculated ranges of $2.40$--$4.26\AU$ and $1.90$--$4.50\AU$
for the conservative and optimistic HZ boundaries, respectively. The
outer companion, with a semi-major axis $3.70\pm0.03\AU$, periastron
$2.44\pm0.03\AU$ and apastron $4.96\pm0.05\AU$, spends most of its
orbit in the HZ by either definition, and might host habitable moons
\citep{hinkel2013,hill2018}.

The strong degeneracies between age, mass and luminosity make brown
dwarfs with independent age estimates invaluable benchmarks for
testing models of substellar evolution
\citep[e.g.][]{marley2015,bowler2016}.  While the expected separation
($\sim70\,\mathrm{mas}$) and contrast ($\sim10^{-7}$) between
\thisstar{} and its brown dwarf companion are beyond the capabilities
of current adaptive optics instruments to measure the brown dwarf's
luminosity and thus test stellar models directly, we can use the
asteroseismic age of the primary to constrain its expected properties.
For example, linearly interpolating the models by \citet{baraffe2003}
using the mass reported by \citet{luhn2019}, increased by 3.2 per cent
to account for our higher estimate of the star's mass, and our age
constraint of $\tfinal{}$ yields $\Teff\approx 560\K$,
$R\approx0.985\RJ$ and $\log_{10}(L/\Lsun)\approx-6.13$, consistent
with a Y-dwarf near the planetary mass boundary.

\section{Conclusions}

We have measured robust asteroseismic properties for the planet host
\thisstar{} by analysing its solar-like oscillations from TESS and
complementary non-seismic parameters with five different stellar
modelling pipelines.  We infer a stellar mass
$M=\Mfinal{}$, radius $R=\Rfinal{}$ and age $t=\tfinal{}$.  Our mass
measurement is near the upper end of the range that has appeared in
the literature.  Our radius measurement is consistent with the Gaia
DR2 and previous
interferometric values, when combined with the new Gaia parallax
measurement.

It is unclear how much more can be extracted from the
asteroseismology of \thisstar{}.  Though TESS will observe the
Southern hemisphere again in its Cycle 3, \thisstar{} will narrowly
miss being re-observed, falling in the gap between Sectors 32 and 33
according to the currently planned satellite pointings.
A more advanced reduction of the existing photometry, however,
might raise several more oscillation modes above the noise level.
Five additional oscillations modes in Table~\ref{t:seis} were identified by two
of the three methods.  If these were all robustly detected, the
substantial increase in seismic data could warrant a new analysis that
would yield a more detailed picture of the star's properties.

Nevertheless, our results demonstrate that precise stellar parameters
can be recovered from relatively poor asteroseismic observations.
Despite measuring only eight oscillation mode frequencies, we have
measured the mass and radius to within $2.7$ and $1.1$ per cent, which
are within the limits of 2 and 15 per cent required for PLATO's core
scientific objectives \citep{goupil2017}.
Our age estimate is slightly less precise ($13.6$ per
cent) than PLATO's requirement of 10 per cent for main-sequence stars.
The longer duration of
PLATO's observations should provide more precise frequency estimates,
even in cases where few modes are detected, so our results suggest
that PLATO's requirements can be met in relatively faint subgiants
($G\approx11$).  Above all, our results imply that TESS has itself
observed many more stars that are interesting (aside from their
oscillations) and could be analysed asteroseismically, even if the
seismic data appears poor.

\section*{Acknowledgements}

WHB, WJC and MBN thank the UK Science and Technology Facilities Council
(STFC) for support under grant ST/R0023297/1.
Funding for the Stellar Astrophysics Centre is provided by The
Danish National Research Foundation (Grant agreement no.: DNRF106).
LGC thanks the support from grant FPI-SO from the Spanish Ministry of Economy and Competitiveness (MINECO) (research project SEV-2015-0548-17-2
and predoctoral contract BES-2017-082610).
SM acknowledges support from the Spanish Ministry with the Ramon y Cajal fellowship number RYC-2015-17697.
ARGS acknowledges the support from NASA under Grant No. NNX17AF27G.
RAG acknowledges the support of the PLATO-CNES grant.
DLB acknowledges support from the TESS GI Program under NASA awards 80NSSC18K1585 and 80NSSC19K0385.
JRM acknowledges support from the Carlsberg Foundation (grant
agreement CF19-0649).
VSA acknowledges support from the Independent Research Fund Denmark
(Research grant 7027-00096B).
BN acknowledges postdoctoral funding from the Alexander von Humboldt Foundation taken at the Max-Planck-Institut f\"{u}r Astrophysik (MPA).
MSC and MD are supported in the form of work contracts funded by
national funds through Funda{\c c}{\~a}o para a Ci{\^e}ncia e Tecnologia
(FCT). MSC and MD acknowledge support by FCT/MCTES through national
funds (PIDDAC) by grants UIDB/04434/2020, UIDP/04434/2020 and
PTDC/FIS-AST/30389/2017 and by FEDER (Fundo Europeu de
Desenvolvimento Regional) through COMPETE2020: Programa Operacional
Competitividade e Internacionaliza{\c c}{\~a}o by grant
POCI-01-0145-FEDER-030389.
TC acknowledges support from the European Union's Horizon 2020 research and innovation programme under the Marie Sk\l{}odowska-Curie grant agreement No.~792848 (PULSATION).
SB acknowledges NASA grants NNX16AI09G and 80NSSC19K0374.
ZCO, MY and S{\"O} acknowledge the Scientific and Technological
Research Council of Turkey (T{\"U}B{\.I}TAK:118F352)
This paper includes data collected by the TESS mission, which are
publicly available from the Mikulski Archive for Space Telescopes
(MAST).  Funding for the TESS mission is provided by the NASA Explorer
Program.  Calculations in this paper made use of the University of
Birmingham's BlueBEAR High-Performance Computing
service.\footnote{\url{http://www.birmingham.ac.uk/bear}}

\section*{Data availability}

Original TESS lightcurves and pixel-level data are available from the
Mikulski Archive for Space Telescopes at \url{http://mast.stsci.edu/}.
Other data underlying this article will be shared on reasonable
request to the corresponding author.

\bibliographystyle{mnras}

\begin{thebibliography}{}
\makeatletter
\relax
\def\mn@urlcharsother{\let\do\@makeother \do\$\do\&\do\#\do\^\do\_\do\%\do\~}
\def\mn@doi{\begingroup\mn@urlcharsother \@ifnextchar [ {\mn@doi@}
  {\mn@doi@[]}}
\def\mn@doi@[#1]#2{\def\@tempa{#1}\ifx\@tempa\@empty \href
  {http://dx.doi.org/#2} {doi:#2}\else \href {http://dx.doi.org/#2} {#1}\fi
  \endgroup}
\def\mn@eprint#1#2{\mn@eprint@#1:#2::\@nil}
\def\mn@eprint@arXiv#1{\href {http://arxiv.org/abs/#1} {{\tt arXiv:#1}}}
\def\mn@eprint@dblp#1{\href {http://dblp.uni-trier.de/rec/bibtex/#1.xml}
  {dblp:#1}}
\def\mn@eprint@#1:#2:#3:#4\@nil{\def\@tempa {#1}\def\@tempb {#2}\def\@tempc
  {#3}\ifx \@tempc \@empty \let \@tempc \@tempb \let \@tempb \@tempa \fi \ifx
  \@tempb \@empty \def\@tempb {arXiv}\fi \@ifundefined
  {mn@eprint@\@tempb}{\@tempb:\@tempc}{\expandafter \expandafter \csname
  mn@eprint@\@tempb\endcsname \expandafter{\@tempc}}}

\bibitem[\protect\citeauthoryear{{Adelberger} et~al.,}{{Adelberger}
  et~al.}{1998}]{adelberger1998}
{Adelberger} E.~G.,  et~al., 1998, \mn@doi [Reviews of Modern Physics]
  {10.1103/RevModPhys.70.1265}, \href
  {http://adsabs.harvard.edu/abs/1998RvMP...70.1265A} {70, 1265}

\bibitem[\protect\citeauthoryear{{Aerts}, {Christensen-Dalsgaard}  \&
  {Kurtz}}{{Aerts} et~al.}{2010}]{acdk2010}
{Aerts} C.,  {Christensen-Dalsgaard} J.,   {Kurtz} D.~W.,  2010,
  {Asteroseismology}.
Astronomy and Astrophysics Library, Springer, Berlin

\bibitem[\protect\citeauthoryear{{Allende Prieto} \& {Lambert}}{{Allende
  Prieto} \& {Lambert}}{1999}]{allendeprieto1999}
{Allende Prieto} C.,  {Lambert} D.~L.,  1999, \aap, \href
  {https://ui.adsabs.harvard.edu/abs/1999A&A...352..555A} {352, 555}

\bibitem[\protect\citeauthoryear{{Andrae} et~al.,}{{Andrae}
  et~al.}{2018}]{andrae2018}
{Andrae} R.,  et~al., 2018, \mn@doi [\aap] {10.1051/0004-6361/201732516}, \href
  {https://ui.adsabs.harvard.edu/abs/2018A&A...616A...8A} {616, A8}

\bibitem[\protect\citeauthoryear{{Angulo} et~al.,}{{Angulo}
  et~al.}{1999}]{angulo1999}
{Angulo} C.,  et~al., 1999, \mn@doi [Nuclear Physics A]
  {10.1016/S0375-9474(99)00030-5}, \href
  {http://adsabs.harvard.edu/abs/1999NuPhA.656....3A} {656, 3}

\bibitem[\protect\citeauthoryear{{Antia} \& {Basu}}{{Antia} \&
  {Basu}}{1994}]{antia1994}
{Antia} H.~M.,  {Basu} S.,  1994, \aaps, \href
  {http://adsabs.harvard.edu/abs/1994A%26AS..107..421A} {107, 421}

\bibitem[\protect\citeauthoryear{{Asplund}, {Grevesse}, {Sauval}  \&
  {Scott}}{{Asplund} et~al.}{2009}]{asplund2009}
{Asplund} M.,  {Grevesse} N.,  {Sauval} A.~J.,   {Scott} P.,  2009, \mn@doi
  [\araa] {10.1146/annurev.astro.46.060407.145222}, \href
  {http://esoads.eso.org/abs/2009ARA%26A..47..481A} {47, 481}

\bibitem[\protect\citeauthoryear{{Baglin}, {Auvergne}, {Barge}, {Deleuil},
  {Catala}, {Michel}, {Weiss}  \& {COROT Team}}{{Baglin}
  et~al.}{2006}]{baglin2006}
{Baglin} A.,  {Auvergne} M.,  {Barge} P.,  {Deleuil} M.,  {Catala} C.,
  {Michel} E.,  {Weiss} W.,   {COROT Team} 2006, in {Fridlund} M.,  {Baglin}
  A.,  {Lochard} J.,   {Conroy} L.,  eds,  ESA Special Publication Vol. 1306,
  ESA Special Publication. p.~33

\bibitem[\protect\citeauthoryear{{Baines}, {McAlister}, {ten Brummelaar},
  {Turner}, {Sturmann}, {Sturmann}, {Goldfinger}  \& {Ridgway}}{{Baines}
  et~al.}{2008}]{baines2008a}
{Baines} E.~K.,  {McAlister} H.~A.,  {ten Brummelaar} T.~A.,  {Turner} N.~H.,
  {Sturmann} J.,  {Sturmann} L.,  {Goldfinger} P.~J.,   {Ridgway} S.~T.,  2008,
  \mn@doi [\apj] {10.1086/588009}, \href
  {https://ui.adsabs.harvard.edu/abs/2008ApJ...680..728B} {680, 728}

\bibitem[\protect\citeauthoryear{{Ball}}{{Ball}}{2017}]{ball2017kasc}
{Ball} W.~H.,  2017, in European Physical Journal Web of Conferences. p. 02001
  (\mn@eprint {arXiv} {1711.01271}), \mn@doi{10.1051/epjconf/201716002001}

\bibitem[\protect\citeauthoryear{{Ball} \& {Gizon}}{{Ball} \&
  {Gizon}}{2014}]{ball2014}
{Ball} W.~H.,  {Gizon} L.,  2014, \mn@doi [\aap] {10.1051/0004-6361/201424325},
  \href {http://adsabs.harvard.edu/abs/2014A%26A...568A.123B} {568, A123}

\bibitem[\protect\citeauthoryear{{Ball} \& {Gizon}}{{Ball} \&
  {Gizon}}{2017}]{ball2017}
{Ball} W.~H.,  {Gizon} L.,  2017, \mn@doi [\aap] {10.1051/0004-6361/201630260},
  \href {http://ukads.nottingham.ac.uk/abs/2017A%26A...600A.128B} {600, A128}

\bibitem[\protect\citeauthoryear{{Baraffe}, {Chabrier}, {Barman}, {Allard}  \&
  {Hauschildt}}{{Baraffe} et~al.}{2003}]{baraffe2003}
{Baraffe} I.,  {Chabrier} G.,  {Barman} T.~S.,  {Allard} F.,   {Hauschildt}
  P.~H.,  2003, \mn@doi [\aap] {10.1051/0004-6361:20030252}, \href
  {https://ui.adsabs.harvard.edu/abs/2003A&A...402..701B} {402, 701}

\bibitem[\protect\citeauthoryear{{Benedict}, {McArthur}, {Bean}, {Barnes},
  {Harrison}, {Hatzes}, {Martioli}  \& {Nelan}}{{Benedict}
  et~al.}{2010}]{benedict2010}
{Benedict} G.~F.,  {McArthur} B.~E.,  {Bean} J.~L.,  {Barnes} R.,  {Harrison}
  T.~E.,  {Hatzes} A.,  {Martioli} E.,   {Nelan} E.~P.,  2010, \mn@doi [\aj]
  {10.1088/0004-6256/139/5/1844}, \href
  {https://ui.adsabs.harvard.edu/abs/2010AJ....139.1844B} {139, 1844}

\bibitem[\protect\citeauthoryear{{Bonfanti}, {Ortolani}  \&
  {Nascimbeni}}{{Bonfanti} et~al.}{2016}]{bonfanti2016}
{Bonfanti} A.,  {Ortolani} S.,   {Nascimbeni} V.,  2016, \mn@doi [\aap]
  {10.1051/0004-6361/201527297}, \href
  {https://ui.adsabs.harvard.edu/abs/2016A&A...585A...5B} {585, A5}

\bibitem[\protect\citeauthoryear{{Borucki} et~al.,}{{Borucki}
  et~al.}{2010}]{kepler}
{Borucki} W.~J.,  et~al., 2010, \mn@doi [Science] {10.1126/science.1185402},
  \href {http://adsabs.harvard.edu/abs/2010Sci...327..977B} {327, 977}

\bibitem[\protect\citeauthoryear{{Bowler}}{{Bowler}}{2016}]{bowler2016}
{Bowler} B.~P.,  2016, \mn@doi [\pasp] {10.1088/1538-3873/128/968/102001},
  \href {https://ui.adsabs.harvard.edu/abs/2016PASP..128j2001B} {128, 102001}

\bibitem[\protect\citeauthoryear{{Brewer}, {Fischer}, {Valenti}  \&
  {Piskunov}}{{Brewer} et~al.}{2016}]{brewer2016}
{Brewer} J.~M.,  {Fischer} D.~A.,  {Valenti} J.~A.,   {Piskunov} N.,  2016,
  \mn@doi [\apjs] {10.3847/0067-0049/225/2/32}, \href
  {https://ui.adsabs.harvard.edu/abs/2016ApJS..225...32B} {225, 32}

\bibitem[\protect\citeauthoryear{{Campante} et~al.,}{{Campante}
  et~al.}{2019}]{campante2019}
{Campante} T.~L.,  et~al., 2019, \mn@doi [\apj] {10.3847/1538-4357/ab44a8},
  \href {https://ui.adsabs.harvard.edu/abs/2019ApJ...885...31C} {885, 31}

\bibitem[\protect\citeauthoryear{{Chen} \& {Zhao}}{{Chen} \&
  {Zhao}}{2002}]{chen2002}
{Chen} Y.-Q.,  {Zhao} G.,  2002, \mn@doi [\cjaa] {10.1088/1009-9271/2/2/151},
  \href {https://ui.adsabs.harvard.edu/abs/2002ChJAA...2..151C} {2, 151}

\bibitem[\protect\citeauthoryear{{Christensen-Dalsgaard}}{{Christensen-Dalsgaard}}{2008}]{adipls}
{Christensen-Dalsgaard} J.,  2008, \mn@doi [\apss] {10.1007/s10509-007-9689-z},
  \href {http://adsabs.harvard.edu/abs/2008Ap%26SS.316..113C} {316, 113}

\bibitem[\protect\citeauthoryear{{CoRot Team}}{{CoRot Team}}{2016}]{corot2016}
{CoRot Team} 2016, {The CoRoT Legacy Book: The adventure of the ultra high
  precision photometry from space, by the CoRot Team}.
EDP Sciences, \mn@doi{10.1051/978-2-7598-1876-1}

\bibitem[\protect\citeauthoryear{{Deal}, {Alecian}, {Lebreton}, {Goupil},
  {Marques}, {LeBlanc}, {Morel}  \& {Pichon}}{{Deal} et~al.}{2018}]{deal2018}
{Deal} M.,  {Alecian} G.,  {Lebreton} Y.,  {Goupil} M.~J.,  {Marques} J.~P.,
  {LeBlanc} F.,  {Morel} P.,   {Pichon} B.,  2018, \mn@doi [\aap]
  {10.1051/0004-6361/201833361}, \href
  {https://ui.adsabs.harvard.edu/abs/2018A&A...618A..10D} {618, A10}

\bibitem[\protect\citeauthoryear{{Deal}, {Goupil}, {Marques}, {Reese}  \&
  {Lebreton}}{{Deal} et~al.}{2020}]{deal2020}
{Deal} M.,  {Goupil} M.~J.,  {Marques} J.~P.,  {Reese} D.~R.,   {Lebreton} Y.,
  2020, \mn@doi [\aap] {10.1051/0004-6361/201936666}, \href
  {https://ui.adsabs.harvard.edu/abs/2020A&A...633A..23D} {633, A23}

\bibitem[\protect\citeauthoryear{{Deheuvels} \& {Michel}}{{Deheuvels} \&
  {Michel}}{2011}]{deheuvels2011}
{Deheuvels} S.,  {Michel} E.,  2011, \mn@doi [\aap]
  {10.1051/0004-6361/201117232}, \href
  {https://ui.adsabs.harvard.edu/abs/2011A&A...535A..91D} {535, A91}

\bibitem[\protect\citeauthoryear{{Deheuvels} et~al.,}{{Deheuvels}
  et~al.}{2014}]{deheuvels2014}
{Deheuvels} S.,  et~al., 2014, \mn@doi [\aap] {10.1051/0004-6361/201322779},
  \href {http://esoads.eso.org/abs/2014A%26A...564A..27D} {564, A27}

\bibitem[\protect\citeauthoryear{{Deka-Szymankiewicz}, {Niedzielski},
  {Adamczyk}, {Adam{\'o}w}, {Nowak}  \& {Wolszczan}}{{Deka-Szymankiewicz}
  et~al.}{2018}]{deka2018}
{Deka-Szymankiewicz} B.,  {Niedzielski} A.,  {Adamczyk} M.,  {Adam{\'o}w} M.,
  {Nowak} G.,   {Wolszczan} A.,  2018, \mn@doi [\aap]
  {10.1051/0004-6361/201731696}, \href
  {https://ui.adsabs.harvard.edu/abs/2018A&A...615A..31D} {615, A31}

\bibitem[\protect\citeauthoryear{{Eggenberger}, {Montalb{\'a}n}  \&
  {Miglio}}{{Eggenberger} et~al.}{2012}]{eggenberger2012}
{Eggenberger} P.,  {Montalb{\'a}n} J.,   {Miglio} A.,  2012, \mn@doi [\aap]
  {10.1051/0004-6361/201219729}, \href
  {http://ukads.nottingham.ac.uk/abs/2012A%26A...544L...4E} {544, L4}

\bibitem[\protect\citeauthoryear{{Ferguson}, {Alexander}, {Allard}, {Barman},
  {Bodnarik}, {Hauschildt}, {Heffner-Wong}  \& {Tamanai}}{{Ferguson}
  et~al.}{2005}]{ferguson2005}
{Ferguson} J.~W.,  {Alexander} D.~R.,  {Allard} F.,  {Barman} T.,  {Bodnarik}
  J.~G.,  {Hauschildt} P.~H.,  {Heffner-Wong} A.,   {Tamanai} A.,  2005,
  \mn@doi [\apj] {10.1086/428642}, \href
  {http://adsabs.harvard.edu/abs/2005ApJ...623..585F} {623, 585}

\bibitem[\protect\citeauthoryear{{Fischer} \& {Valenti}}{{Fischer} \&
  {Valenti}}{2005}]{fischer2005}
{Fischer} D.~A.,  {Valenti} J.,  2005, \mn@doi [\apj] {10.1086/428383}, \href
  {https://ui.adsabs.harvard.edu/abs/2005ApJ...622.1102F} {622, 1102}

\bibitem[\protect\citeauthoryear{{Fischer}, {Marcy}, {Butler}, {Vogt}, {Frink}
  \& {Apps}}{{Fischer} et~al.}{2001}]{fischer2001}
{Fischer} D.~A.,  {Marcy} G.~W.,  {Butler} R.~P.,  {Vogt} S.~S.,  {Frink} S.,
  {Apps} K.,  2001, \mn@doi [\apj] {10.1086/320224}, \href
  {https://ui.adsabs.harvard.edu/abs/2001ApJ...551.1107F} {551, 1107}

\bibitem[\protect\citeauthoryear{{Fischer} et~al.,}{{Fischer}
  et~al.}{2003}]{fischer2003}
{Fischer} D.~A.,  et~al., 2003, \mn@doi [\apj] {10.1086/367889}, \href
  {https://ui.adsabs.harvard.edu/abs/2003ApJ...586.1394F} {586, 1394}

\bibitem[\protect\citeauthoryear{{Formicola} et~al.,}{{Formicola}
  et~al.}{2004}]{formicola2004}
{Formicola} A.,  et~al., 2004, \mn@doi [Physics Letters B]
  {10.1016/j.physletb.2004.03.092}, \href
  {https://ui.adsabs.harvard.edu/abs/2004PhLB..591...61F} {591, 61}

\bibitem[\protect\citeauthoryear{{Fuhrmann}}{{Fuhrmann}}{2008}]{fuhrmann2008}
{Fuhrmann} K.,  2008, \mn@doi [\mnras] {10.1111/j.1365-2966.2007.12671.x},
  \href {https://ui.adsabs.harvard.edu/abs/2008MNRAS.384..173F} {384, 173}

\bibitem[\protect\citeauthoryear{{Garc{\'\i}a} \& {Ballot}}{{Garc{\'\i}a} \&
  {Ballot}}{2019}]{garcia2019}
{Garc{\'\i}a} R.~A.,  {Ballot} J.,  2019, \mn@doi [Living Reviews in Solar
  Physics] {10.1007/s41116-019-0020-1}, \href
  {https://ui.adsabs.harvard.edu/abs/2019LRSP...16....4G} {16, 4}

\bibitem[\protect\citeauthoryear{{Garc{\'\i}a} et~al.,}{{Garc{\'\i}a}
  et~al.}{2011}]{garcia2011}
{Garc{\'\i}a} R.~A.,  et~al., 2011, \mn@doi [\mnras]
  {10.1111/j.1745-3933.2011.01042.x}, \href
  {https://ui.adsabs.harvard.edu/abs/2011MNRAS.414L...6G} {414, L6}

\bibitem[\protect\citeauthoryear{{Garc{\'\i}a} et~al.,}{{Garc{\'\i}a}
  et~al.}{2014}]{garcia2014}
{Garc{\'\i}a} R.~A.,  et~al., 2014, \mn@doi [\aap]
  {10.1051/0004-6361/201323326}, \href
  {https://ui.adsabs.harvard.edu/abs/2014A&A...568A..10G} {568, A10}

\bibitem[\protect\citeauthoryear{{Ghezzi}, {Cunha}, {Schuler}  \&
  {Smith}}{{Ghezzi} et~al.}{2010}]{ghezzi2010}
{Ghezzi} L.,  {Cunha} K.,  {Schuler} S.~C.,   {Smith} V.~V.,  2010, \mn@doi
  [\apj] {10.1088/0004-637X/725/1/721}, \href
  {https://ui.adsabs.harvard.edu/abs/2010ApJ...725..721G} {725, 721}

\bibitem[\protect\citeauthoryear{{Gonzalez}, {Laws}, {Tyagi}  \&
  {Reddy}}{{Gonzalez} et~al.}{2001}]{gonzalez2001}
{Gonzalez} G.,  {Laws} C.,  {Tyagi} S.,   {Reddy} B.~E.,  2001, \mn@doi [\aj]
  {10.1086/318048}, \href
  {https://ui.adsabs.harvard.edu/abs/2001AJ....121..432G} {121, 432}

\bibitem[\protect\citeauthoryear{{Goupil}}{{Goupil}}{2017}]{goupil2017}
{Goupil} M.,  2017, in European Physical Journal Web of Conferences. p. 01003,
  \mn@doi{10.1051/epjconf/201716001003}

\bibitem[\protect\citeauthoryear{{Grevesse} \& {Noels}}{{Grevesse} \&
  {Noels}}{1993}]{grevesse1993}
{Grevesse} N.,  {Noels} A.,  1993, in {Prantzos} N.,  {Vangioni-Flam} E.,
  {Casse} M.,  eds, Origin and Evolution of the Elements. pp 15--25

\bibitem[\protect\citeauthoryear{{Grevesse} \& {Sauval}}{{Grevesse} \&
  {Sauval}}{1998}]{grevesse1998}
{Grevesse} N.,  {Sauval} A.~J.,  1998, \mn@doi [\ssr]
  {10.1023/A:1005161325181}, \href
  {http://adsabs.harvard.edu/abs/1998SSRv...85..161G} {85, 161}

\bibitem[\protect\citeauthoryear{{Hammer} et~al.,}{{Hammer}
  et~al.}{2005}]{hammer2005}
{Hammer} J.~W.,  et~al., 2005, \mn@doi [\nphysa]
  {10.1016/j.nuclphysa.2005.05.066}, \href
  {https://ui.adsabs.harvard.edu/abs/2005NuPhA.758..363H} {758, 363}

\bibitem[\protect\citeauthoryear{{Hekker} \& {Christensen-Dalsgaard}}{{Hekker}
  \& {Christensen-Dalsgaard}}{2017}]{hekker2017}
{Hekker} S.,  {Christensen-Dalsgaard} J.,  2017, \mn@doi [\aapr]
  {10.1007/s00159-017-0101-x}, \href
  {https://ui.adsabs.harvard.edu/abs/2017A&ARv..25....1H} {25, 1}

\bibitem[\protect\citeauthoryear{{Henry} et~al.,}{{Henry}
  et~al.}{2013}]{henry2013}
{Henry} G.~W.,  et~al., 2013, \mn@doi [\apj] {10.1088/0004-637X/768/2/155},
  \href {https://ui.adsabs.harvard.edu/abs/2013ApJ...768..155H} {768, 155}

\bibitem[\protect\citeauthoryear{{Hill}, {Kane}, {Seperuelo Duarte},
  {Kopparapu}, {Gelino}  \& {Wittenmyer}}{{Hill} et~al.}{2018}]{hill2018}
{Hill} M.~L.,  {Kane} S.~R.,  {Seperuelo Duarte} E.,  {Kopparapu} R.~K.,
  {Gelino} D.~M.,   {Wittenmyer} R.~A.,  2018, \mn@doi [\apj]
  {10.3847/1538-4357/aac384}, \href
  {https://ui.adsabs.harvard.edu/abs/2018ApJ...860...67H} {860, 67}

\bibitem[\protect\citeauthoryear{{Hinkel} \& {Kane}}{{Hinkel} \&
  {Kane}}{2013}]{hinkel2013}
{Hinkel} N.~R.,  {Kane} S.~R.,  2013, \mn@doi [\apj]
  {10.1088/0004-637X/774/1/27}, \href
  {https://ui.adsabs.harvard.edu/abs/2013ApJ...774...27H} {774, 27}

\bibitem[\protect\citeauthoryear{{H{\o}g} et~al.,}{{H{\o}g}
  et~al.}{2000a}]{tycho2a}
{H{\o}g} E.,  et~al., 2000a, \aap, \href
  {https://ui.adsabs.harvard.edu/abs/2000A&A...355L..27H} {355, L27}

\bibitem[\protect\citeauthoryear{{H{\o}g} et~al.,}{{H{\o}g}
  et~al.}{2000b}]{tycho2b}
{H{\o}g} E.,  et~al., 2000b, \aap, \href
  {https://ui.adsabs.harvard.edu/abs/2000A&A...357..367H} {357, 367}

\bibitem[\protect\citeauthoryear{{Howell} et~al.,}{{Howell} et~al.}{2014}]{k2}
{Howell} S.~B.,  et~al., 2014, \mn@doi [\pasp] {10.1086/676406}, \href
  {http://adsabs.harvard.edu/abs/2014PASP..126..398H} {126, 398}

\bibitem[\protect\citeauthoryear{{Iglesias} \& {Rogers}}{{Iglesias} \&
  {Rogers}}{1993}]{iglesias1993}
{Iglesias} C.~A.,  {Rogers} F.~J.,  1993, \mn@doi [\apj] {10.1086/172958},
  \href {https://ui.adsabs.harvard.edu/abs/1993ApJ...412..752I} {412, 752}

\bibitem[\protect\citeauthoryear{{Iglesias} \& {Rogers}}{{Iglesias} \&
  {Rogers}}{1996}]{iglesias1996}
{Iglesias} C.~A.,  {Rogers} F.~J.,  1996, \mn@doi [\apj] {10.1086/177381},
  \href {http://adsabs.harvard.edu/abs/1996ApJ...464..943I} {464, 943}

\bibitem[\protect\citeauthoryear{{Imbriani} et~al.,}{{Imbriani}
  et~al.}{2005}]{imbriani2005}
{Imbriani} G.,  et~al., 2005, \mn@doi [European Physical Journal A]
  {10.1140/epja/i2005-10138-7}, \href
  {http://esoads.eso.org/abs/2005EPJA...25..455I} {25, 455}

\bibitem[\protect\citeauthoryear{{Jenkins} et~al.,}{{Jenkins}
  et~al.}{2016}]{spoc}
{Jenkins} J.~M.,  et~al., 2016, in Chiozzi G.,  Guzman J.~C.,  eds,  Society of
  Photo-Optical Instrumentation Engineers (SPIE) Conference Series Vol. 9913,
  Software and Cyberinfrastructure for Astronomy IV. SPIE, pp 1232 -- 1251,
  \mn@doi{10.1117/12.2233418}, \url {https://doi.org/10.1117/12.2233418}

\bibitem[\protect\citeauthoryear{{Jofr{\'e}}, {Petrucci}, {Saffe}, {Saker},
  {Artur de la Villarmois}, {Chavero}, {G{\'o}mez}  \& {Mauas}}{{Jofr{\'e}}
  et~al.}{2015}]{jofre2015}
{Jofr{\'e}} E.,  {Petrucci} R.,  {Saffe} C.,  {Saker} L.,  {Artur de la
  Villarmois} E.,  {Chavero} C.,  {G{\'o}mez} M.,   {Mauas} P.~J.~D.,  2015,
  \mn@doi [\aap] {10.1051/0004-6361/201424474}, \href
  {https://ui.adsabs.harvard.edu/abs/2015A&A...574A..50J} {574, A50}

\bibitem[\protect\citeauthoryear{{Kane}}{{Kane}}{2014}]{kane2014}
{Kane} S.~R.,  2014, \mn@doi [\apj] {10.1088/0004-637X/782/2/111}, \href
  {https://ui.adsabs.harvard.edu/abs/2014ApJ...782..111K} {782, 111}

\bibitem[\protect\citeauthoryear{{Kane}, {Mahadevan}, {von Braun}, {Laughlin}
  \& {Ciardi}}{{Kane} et~al.}{2009}]{kane2009}
{Kane} S.~R.,  {Mahadevan} S.,  {von Braun} K.,  {Laughlin} G.,   {Ciardi}
  D.~R.,  2009, \mn@doi [\pasp] {10.1086/648564}, \href
  {https://ui.adsabs.harvard.edu/abs/2009PASP..121.1386K} {121, 1386}

\bibitem[\protect\citeauthoryear{{Kane} et~al.,}{{Kane}
  et~al.}{2016}]{kane2016}
{Kane} S.~R.,  et~al., 2016, \mn@doi [\apj] {10.3847/0004-637X/830/1/1}, \href
  {https://ui.adsabs.harvard.edu/abs/2016ApJ...830....1K} {830, 1}

\bibitem[\protect\citeauthoryear{{Kang}, {Lee}  \& {Kim}}{{Kang}
  et~al.}{2011}]{kang2011}
{Kang} W.,  {Lee} S.-G.,   {Kim} K.-M.,  2011, \mn@doi [\apj]
  {10.1088/0004-637X/736/2/87}, \href
  {https://ui.adsabs.harvard.edu/abs/2011ApJ...736...87K} {736, 87}

\bibitem[\protect\citeauthoryear{{Kasting}, {Whitmire}  \&
  {Reynolds}}{{Kasting} et~al.}{1993}]{kasting1993}
{Kasting} J.~F.,  {Whitmire} D.~P.,   {Reynolds} R.~T.,  1993, \mn@doi
  [\icarus] {10.1006/icar.1993.1010}, \href
  {https://ui.adsabs.harvard.edu/abs/1993Icar..101..108K} {101, 108}

\bibitem[\protect\citeauthoryear{{Kjeldsen}, {Bedding}  \&
  {Christensen-Dalsgaard}}{{Kjeldsen} et~al.}{2008}]{kjeldsen2008}
{Kjeldsen} H.,  {Bedding} T.~R.,   {Christensen-Dalsgaard} J.,  2008, \mn@doi
  [\apjl] {10.1086/591667}, \href
  {http://esoads.eso.org/abs/2008ApJ...683L.175K} {683, L175}

\bibitem[\protect\citeauthoryear{{Kopparapu} et~al.,}{{Kopparapu}
  et~al.}{2013}]{kopparapu2013}
{Kopparapu} R.~K.,  et~al., 2013, \mn@doi [\apj] {10.1088/0004-637X/765/2/131},
  \href {https://ui.adsabs.harvard.edu/abs/2013ApJ...765..131K} {765, 131}

\bibitem[\protect\citeauthoryear{{Kopparapu}, {Ramirez}, {SchottelKotte},
  {Kasting}, {Domagal-Goldman}  \& {Eymet}}{{Kopparapu}
  et~al.}{2014}]{kopparapu2014}
{Kopparapu} R.~K.,  {Ramirez} R.~M.,  {SchottelKotte} J.,  {Kasting} J.~F.,
  {Domagal-Goldman} S.,   {Eymet} V.,  2014, \mn@doi [\apjl]
  {10.1088/2041-8205/787/2/L29}, \href
  {https://ui.adsabs.harvard.edu/abs/2014ApJ...787L..29K} {787, L29}

\bibitem[\protect\citeauthoryear{{Kunz}, {Fey}, {Jaeger}, {Mayer}, {Hammer},
  {Staudt}, {Harissopulos}  \& {Paradellis}}{{Kunz} et~al.}{2002}]{kunz2002}
{Kunz} R.,  {Fey} M.,  {Jaeger} M.,  {Mayer} A.,  {Hammer} J.~W.,  {Staudt} G.,
   {Harissopulos} S.,   {Paradellis} T.,  2002, \mn@doi [\apj]
  {10.1086/338384}, \href {http://adsabs.harvard.edu/abs/2002ApJ...567..643K}
  {567, 643}

\bibitem[\protect\citeauthoryear{{Kurucz}}{{Kurucz}}{2013}]{kurucz2013}
{Kurucz} R.~L.,  2013, {ATLAS12: Opacity sampling model atmosphere program}
  (\mn@eprint {ascl} {1303.024})

\bibitem[\protect\citeauthoryear{{Laws}, {Gonzalez}, {Walker}, {Tyagi},
  {Dodsworth}, {Snider}  \& {Suntzeff}}{{Laws} et~al.}{2003}]{laws2003}
{Laws} C.,  {Gonzalez} G.,  {Walker} K.~M.,  {Tyagi} S.,  {Dodsworth} J.,
  {Snider} K.,   {Suntzeff} N.~B.,  2003, \mn@doi [\aj] {10.1086/374626}, \href
  {https://ui.adsabs.harvard.edu/abs/2003AJ....125.2664L} {125, 2664}

\bibitem[\protect\citeauthoryear{{Li}, {Bedding}, {Christensen-Dalsgaard},
  {Stello}, {Li}  \& {Keen}}{{Li} et~al.}{2020}]{li2020}
{Li} T.,  {Bedding} T.~R.,  {Christensen-Dalsgaard} J.,  {Stello} D.,  {Li} Y.,
    {Keen} M.~A.,  2020, \mn@doi [\mnras] {10.1093/mnras/staa1350}, \href
  {https://ui.adsabs.harvard.edu/abs/2020MNRAS.495.3431L} {495, 3431}

\bibitem[\protect\citeauthoryear{{Luck}}{{Luck}}{2017}]{luck2017}
{Luck} R.~E.,  2017, \mn@doi [\aj] {10.3847/1538-3881/153/1/21}, \href
  {https://ui.adsabs.harvard.edu/abs/2017AJ....153...21L} {153, 21}

\bibitem[\protect\citeauthoryear{{Luck} \& {Heiter}}{{Luck} \&
  {Heiter}}{2006}]{luck2006}
{Luck} R.~E.,  {Heiter} U.,  2006, \mn@doi [\aj] {10.1086/504080}, \href
  {https://ui.adsabs.harvard.edu/abs/2006AJ....131.3069L} {131, 3069}

\bibitem[\protect\citeauthoryear{{Luhn}, {Bastien}, {Wright}, {Johnson},
  {Howard}  \& {Isaacson}}{{Luhn} et~al.}{2019}]{luhn2019}
{Luhn} J.~K.,  {Bastien} F.~A.,  {Wright} J.~T.,  {Johnson} J.~A.,  {Howard}
  A.~W.,   {Isaacson} H.,  2019, \mn@doi [\aj] {10.3847/1538-3881/aaf5d0},
  \href {https://ui.adsabs.harvard.edu/abs/2019AJ....157..149L} {157, 149}

\bibitem[\protect\citeauthoryear{{Maldonado} \& {Villaver}}{{Maldonado} \&
  {Villaver}}{2016}]{maldonado2016}
{Maldonado} J.,  {Villaver} E.,  2016, \mn@doi [\aap]
  {10.1051/0004-6361/201527883}, \href
  {https://ui.adsabs.harvard.edu/abs/2016A&A...588A..98M} {588, A98}

\bibitem[\protect\citeauthoryear{{Maldonado} \& {Villaver}}{{Maldonado} \&
  {Villaver}}{2017}]{maldonado2017}
{Maldonado} J.,  {Villaver} E.,  2017, \mn@doi [\aap]
  {10.1051/0004-6361/201630120}, \href
  {https://ui.adsabs.harvard.edu/abs/2017A&A...602A..38M} {602, A38}

\bibitem[\protect\citeauthoryear{{Maldonado}, {Villaver}  \&
  {Eiroa}}{{Maldonado} et~al.}{2013}]{maldonado2013}
{Maldonado} J.,  {Villaver} E.,   {Eiroa} C.,  2013, \mn@doi [\aap]
  {10.1051/0004-6361/201321082}, \href
  {https://ui.adsabs.harvard.edu/abs/2013A&A...554A..84M} {554, A84}

\bibitem[\protect\citeauthoryear{{Marley} \& {Robinson}}{{Marley} \&
  {Robinson}}{2015}]{marley2015}
{Marley} M.~S.,  {Robinson} T.~D.,  2015, \mn@doi [\araa]
  {10.1146/annurev-astro-082214-122522}, \href
  {https://ui.adsabs.harvard.edu/abs/2015ARA&A..53..279M} {53, 279}

\bibitem[\protect\citeauthoryear{{Marques} et~al.,}{{Marques}
  et~al.}{2013}]{marques2013}
{Marques} J.~P.,  et~al., 2013, \mn@doi [\aap] {10.1051/0004-6361/201220211},
  \href {http://ukads.nottingham.ac.uk/abs/2013A%26A...549A..74M} {549, A74}

\bibitem[\protect\citeauthoryear{{Mermilliod}}{{Mermilliod}}{2006}]{mermilliod2006}
{Mermilliod} J.~C.,  2006, VizieR Online Data Catalog, \href
  {https://ui.adsabs.harvard.edu/abs/2006yCat.2168....0M} {p. II/168}

\bibitem[\protect\citeauthoryear{{Michaud}, {Alecian}  \& {Richer}}{{Michaud}
  et~al.}{2015}]{michaud2015}
{Michaud} G.,  {Alecian} G.,   {Richer} J.,  2015, {Atomic Diffusion in Stars}.
Springer-Verlag, \mn@doi{10.1007/978-3-319-19854-5}

\bibitem[\protect\citeauthoryear{Mier}{Mier}{2017}]{yabox}
Mier P.~R.,  2017, pablormier/yabox: v1.0.3, \mn@doi{10.5281/zenodo.848679},
  \url {https://doi.org/10.5281/zenodo.848679}

\bibitem[\protect\citeauthoryear{{Mortier}, {Santos}, {Sousa}, {Adibekyan},
  {Delgado Mena}, {Tsantaki}, {Israelian}  \& {Mayor}}{{Mortier}
  et~al.}{2013}]{mortier2013}
{Mortier} A.,  {Santos} N.~C.,  {Sousa} S.~G.,  {Adibekyan} V.~Z.,  {Delgado
  Mena} E.,  {Tsantaki} M.,  {Israelian} G.,   {Mayor} M.,  2013, \mn@doi
  [\aap] {10.1051/0004-6361/201321641}, \href
  {https://ui.adsabs.harvard.edu/abs/2013A&A...557A..70M} {557, A70}

\bibitem[\protect\citeauthoryear{{Mosser} et~al.,}{{Mosser}
  et~al.}{2011}]{mosser2011up}
{Mosser} B.,  et~al., 2011, \mn@doi [\aap] {10.1051/0004-6361/201015440}, \href
  {https://ui.adsabs.harvard.edu/abs/2011A&A...525L...9M} {525, L9}

\bibitem[\protect\citeauthoryear{{Mosser}, {Vrard}, {Belkacem}, {Deheuvels}  \&
  {Goupil}}{{Mosser} et~al.}{2015}]{mosser2015}
{Mosser} B.,  {Vrard} M.,  {Belkacem} K.,  {Deheuvels} S.,   {Goupil} M.~J.,
  2015, \mn@doi [\aap] {10.1051/0004-6361/201527075}, \href
  {https://ui.adsabs.harvard.edu/abs/2015A&A...584A..50M} {584, A50}

\bibitem[\protect\citeauthoryear{{Mosumgaard}, {Ball}, {Silva Aguirre}, {Weiss}
   \& {Christensen-Dalsgaard}}{{Mosumgaard} et~al.}{2018}]{mosumgaard2018}
{Mosumgaard} J.~R.,  {Ball} W.~H.,  {Silva Aguirre} V.,  {Weiss} A.,
  {Christensen-Dalsgaard} J.,  2018, \mn@doi [\mnras] {10.1093/mnras/sty1442},
  \href {https://ui.adsabs.harvard.edu/abs/2018MNRAS.478.5650M} {478, 5650}

\bibitem[\protect\citeauthoryear{Nelder \& Mead}{Nelder \&
  Mead}{1965}]{nelder1965}
Nelder J.~A.,  Mead R.,  1965, \mn@doi [The Computer Journal]
  {10.1093/comjnl/7.4.308}, 7, 308

\bibitem[\protect\citeauthoryear{{Niedzielski}, {Deka-Szymankiewicz},
  {Adamczyk}, {Adam{\'o}w}, {Nowak}  \& {Wolszczan}}{{Niedzielski}
  et~al.}{2016}]{niedzielski2016}
{Niedzielski} A.,  {Deka-Szymankiewicz} B.,  {Adamczyk} M.,  {Adam{\'o}w} M.,
  {Nowak} G.,   {Wolszczan} A.,  2016, \mn@doi [\aap]
  {10.1051/0004-6361/201527362}, \href
  {https://ui.adsabs.harvard.edu/abs/2016A&A...585A..73N} {585, A73}

\bibitem[\protect\citeauthoryear{{Nielsen} et~al.,}{{Nielsen}
  et~al.}{2020}]{nielsen2020}
{Nielsen} M.~B.,  et~al., 2020, \aap

\bibitem[\protect\citeauthoryear{{Paunzen}}{{Paunzen}}{2015}]{paunzen2015}
{Paunzen} E.,  2015, \mn@doi [\aap] {10.1051/0004-6361/201526413}, \href
  {https://ui.adsabs.harvard.edu/abs/2015A&A...580A..23P} {580, A23}

\bibitem[\protect\citeauthoryear{{Paxton}, {Bildsten}, {Dotter}, {Herwig},
  {Lesaffre}  \& {Timmes}}{{Paxton} et~al.}{2011}]{paxton2011}
{Paxton} B.,  {Bildsten} L.,  {Dotter} A.,  {Herwig} F.,  {Lesaffre} P.,
  {Timmes} F.,  2011, \mn@doi [\apjs] {10.1088/0067-0049/192/1/3}, \href
  {http://esoads.eso.org/abs/2011ApJS..192....3P} {192, 3}

\bibitem[\protect\citeauthoryear{{Paxton} et~al.,}{{Paxton}
  et~al.}{2013}]{paxton2013}
{Paxton} B.,  et~al., 2013, \mn@doi [\apjs] {10.1088/0067-0049/208/1/4}, \href
  {http://esoads.eso.org/abs/2013ApJS..208....4P} {208, 4}

\bibitem[\protect\citeauthoryear{{Paxton} et~al.,}{{Paxton}
  et~al.}{2015}]{paxton2015}
{Paxton} B.,  et~al., 2015, \mn@doi [\apjs] {10.1088/0067-0049/220/1/15}, \href
  {http://esoads.eso.org/abs/2015ApJS..220...15P} {220, 15}

\bibitem[\protect\citeauthoryear{{Pires}, {Mathur}, {Garc{\'\i}a}, {Ballot},
  {Stello}  \& {Sato}}{{Pires} et~al.}{2015}]{pires2015}
{Pires} S.,  {Mathur} S.,  {Garc{\'\i}a} R.~A.,  {Ballot} J.,  {Stello} D.,
  {Sato} K.,  2015, \mn@doi [\aap] {10.1051/0004-6361/201322361}, \href
  {https://ui.adsabs.harvard.edu/abs/2015A&A...574A..18P} {574, A18}

\bibitem[\protect\citeauthoryear{{Rendle} et~al.,}{{Rendle}
  et~al.}{2019}]{rendle2019}
{Rendle} B.~M.,  et~al., 2019, \mn@doi [\mnras] {10.1093/mnras/stz031}, \href
  {https://ui.adsabs.harvard.edu/abs/2019MNRAS.484..771R} {484, 771}

\bibitem[\protect\citeauthoryear{{Rogers} \& {Nayfonov}}{{Rogers} \&
  {Nayfonov}}{2002}]{rogers2002}
{Rogers} F.~J.,  {Nayfonov} A.,  2002, \mn@doi [\apj] {10.1086/341894}, \href
  {http://esoads.eso.org/abs/2002ApJ...576.1064R} {576, 1064}

\bibitem[\protect\citeauthoryear{{Roxburgh}}{{Roxburgh}}{2015}]{roxburgh2015}
{Roxburgh} I.~W.,  2015, \mn@doi [\aap] {10.1051/0004-6361/201425289}, \href
  {http://esoads.eso.org/abs/2015A%26A...574A..45R} {574, A45}

\bibitem[\protect\citeauthoryear{{Roxburgh}}{{Roxburgh}}{2016}]{roxburgh2016}
{Roxburgh} I.~W.,  2016, \mn@doi [\aap] {10.1051/0004-6361/201526593}, \href
  {http://esoads.eso.org/abs/2016A%26A...585A..63R} {585, A63}

\bibitem[\protect\citeauthoryear{{Roxburgh} \& {Vorontsov}}{{Roxburgh} \&
  {Vorontsov}}{2003}]{roxburgh2003}
{Roxburgh} I.~W.,  {Vorontsov} S.~V.,  2003, \mn@doi [\aap]
  {10.1051/0004-6361:20031318}, \href
  {http://esoads.eso.org/abs/2003A%26A...411..215R} {411, 215}

\bibitem[\protect\citeauthoryear{{Salaris} \& {Cassisi}}{{Salaris} \&
  {Cassisi}}{2017}]{salaris2017}
{Salaris} M.,  {Cassisi} S.,  2017, \mn@doi [Royal Society Open Science]
  {10.1098/rsos.170192}, \href
  {https://ui.adsabs.harvard.edu/abs/2017RSOS....470192S} {4, 170192}

\bibitem[\protect\citeauthoryear{{Santos}, {Israelian}  \& {Mayor}}{{Santos}
  et~al.}{2001}]{santos2001}
{Santos} N.~C.,  {Israelian} G.,   {Mayor} M.,  2001, \mn@doi [\aap]
  {10.1051/0004-6361:20010648}, \href
  {https://ui.adsabs.harvard.edu/abs/2001A&A...373.1019S} {373, 1019}

\bibitem[\protect\citeauthoryear{{Santos}, {Israelian}  \& {Mayor}}{{Santos}
  et~al.}{2004}]{santos2004}
{Santos} N.~C.,  {Israelian} G.,   {Mayor} M.,  2004, \mn@doi [\aap]
  {10.1051/0004-6361:20034469}, \href
  {https://ui.adsabs.harvard.edu/abs/2004A&A...415.1153S} {415, 1153}

\bibitem[\protect\citeauthoryear{{Schofield} et~al.,}{{Schofield}
  et~al.}{2019}]{schofield2019}
{Schofield} M.,  et~al., 2019, \mn@doi [\apjs] {10.3847/1538-4365/ab04f5},
  \href {https://ui.adsabs.harvard.edu/abs/2019ApJS..241...12S} {241, 12}

\bibitem[\protect\citeauthoryear{{Silva Aguirre} et~al.,}{{Silva Aguirre}
  et~al.}{2015}]{silva2015}
{Silva Aguirre} V.,  et~al., 2015, \mn@doi [\mnras] {10.1093/mnras/stv1388},
  \href {https://ui.adsabs.harvard.edu/abs/2015MNRAS.452.2127S} {452, 2127}

\bibitem[\protect\citeauthoryear{{Silva Aguirre} et~al.,}{{Silva Aguirre}
  et~al.}{2017}]{legacy2}
{Silva Aguirre} V.,  et~al., 2017, \mn@doi [\apj]
  {10.3847/1538-4357/835/2/173}, 835, 173

\bibitem[\protect\citeauthoryear{{Sonoi}, {Samadi}, {Belkacem}, {Ludwig},
  {Caffau}  \& {Mosser}}{{Sonoi} et~al.}{2015}]{sonoi2015}
{Sonoi} T.,  {Samadi} R.,  {Belkacem} K.,  {Ludwig} H.-G.,  {Caffau} E.,
  {Mosser} B.,  2015, \mn@doi [\aap] {10.1051/0004-6361/201526838}, \href
  {http://adsabs.harvard.edu/abs/2015A%26A...583A.112S} {583, A112}

\bibitem[\protect\citeauthoryear{{Spada}, {Gellert}, {Arlt}  \&
  {Deheuvels}}{{Spada} et~al.}{2016}]{spada2016}
{Spada} F.,  {Gellert} M.,  {Arlt} R.,   {Deheuvels} S.,  2016, \mn@doi [\aap]
  {10.1051/0004-6361/201527591}, \href
  {https://ui.adsabs.harvard.edu/abs/2016A&A...589A..23S} {589, A23}

\bibitem[\protect\citeauthoryear{{Spiegel}, {Burrows}  \& {Milsom}}{{Spiegel}
  et~al.}{2011}]{spiegel2011}
{Spiegel} D.~S.,  {Burrows} A.,   {Milsom} J.~A.,  2011, \mn@doi [\apj]
  {10.1088/0004-637X/727/1/57}, \href
  {https://ui.adsabs.harvard.edu/abs/2011ApJ...727...57S} {727, 57}

\bibitem[\protect\citeauthoryear{{Stassun} \& {Torres}}{{Stassun} \&
  {Torres}}{2016}]{stassun2016}
{Stassun} K.~G.,  {Torres} G.,  2016, \mn@doi [\aj]
  {10.3847/0004-6256/152/6/180}, \href
  {https://ui.adsabs.harvard.edu/abs/2016AJ....152..180S} {152, 180}

\bibitem[\protect\citeauthoryear{{Stassun}, {Collins}  \& {Gaudi}}{{Stassun}
  et~al.}{2017}]{stassun2017}
{Stassun} K.~G.,  {Collins} K.~A.,   {Gaudi} B.~S.,  2017, \mn@doi [\aj]
  {10.3847/1538-3881/aa5df3}, \href
  {https://ui.adsabs.harvard.edu/abs/2017AJ....153..136S} {153, 136}

\bibitem[\protect\citeauthoryear{{Stassun}, {Corsaro}, {Pepper}  \&
  {Gaudi}}{{Stassun} et~al.}{2018}]{stassun2018}
{Stassun} K.~G.,  {Corsaro} E.,  {Pepper} J.~A.,   {Gaudi} B.~S.,  2018,
  \mn@doi [\aj] {10.3847/1538-3881/aa998a}, \href
  {https://ui.adsabs.harvard.edu/abs/2018AJ....155...22S} {155, 22}

\bibitem[\protect\citeauthoryear{{Stokholm}, {Nissen}, {Silva Aguirre},
  {White}, {Lund}, {Mosumgaard}, {Huber}  \& {Jessen-Hansen}}{{Stokholm}
  et~al.}{2019}]{stokholm2019}
{Stokholm} A.,  {Nissen} P.~E.,  {Silva Aguirre} V.,  {White} T.~R.,  {Lund}
  M.~N.,  {Mosumgaard} J.~R.,  {Huber} D.,   {Jessen-Hansen} J.,  2019, \mn@doi
  [\mnras] {10.1093/mnras/stz2222}, \href
  {https://ui.adsabs.harvard.edu/abs/2019MNRAS.489..928S} {489, 928}

\bibitem[\protect\citeauthoryear{{Takeda}}{{Takeda}}{2007}]{takeda2007a}
{Takeda} Y.,  2007, \mn@doi [\pasj] {10.1093/pasj/59.2.335}, \href
  {https://ui.adsabs.harvard.edu/abs/2007PASJ...59..335T} {59, 335}

\bibitem[\protect\citeauthoryear{{Takeda}, {Sato}, {Kambe}, {Sadakane}  \&
  {Ohkubo}}{{Takeda} et~al.}{2002}]{takeda2002}
{Takeda} Y.,  {Sato} B.,  {Kambe} E.,  {Sadakane} K.,   {Ohkubo} M.,  2002,
  \mn@doi [\pasj] {10.1093/pasj/54.6.1041}, \href
  {https://ui.adsabs.harvard.edu/abs/2002PASJ...54.1041T} {54, 1041}

\bibitem[\protect\citeauthoryear{{Takeda}, {Ohkubo}, {Sato}, {Kambe}  \&
  {Sadakane}}{{Takeda} et~al.}{2005}]{takeda2005}
{Takeda} Y.,  {Ohkubo} M.,  {Sato} B.,  {Kambe} E.,   {Sadakane} K.,  2005,
  \mn@doi [\pasj] {10.1093/pasj/57.1.27}, \href
  {https://ui.adsabs.harvard.edu/abs/2005PASJ...57...27T} {57, 27}

\bibitem[\protect\citeauthoryear{{Takeda}, {Ford}, {Sills}, {Rasio}, {Fischer}
  \& {Valenti}}{{Takeda} et~al.}{2007}]{takeda2007b}
{Takeda} G.,  {Ford} E.~B.,  {Sills} A.,  {Rasio} F.~A.,  {Fischer} D.~A.,
  {Valenti} J.~A.,  2007, \mn@doi [\apjs] {10.1086/509763}, \href
  {https://ui.adsabs.harvard.edu/abs/2007ApJS..168..297T} {168, 297}

\bibitem[\protect\citeauthoryear{{Tasoulis}, {Pavlidis}, {Plagianakos}  \&
  {Vrahatis}}{{Tasoulis} et~al.}{2004}]{tasoulis2004}
{Tasoulis} D.~K.,  {Pavlidis} N.~G.,  {Plagianakos} V.~P.,   {Vrahatis} M.~N.,
  2004, in Proceedings of the 2004 Congress on Evolutionary Computation (IEEE
  Cat. No.04TH8753). pp 2023--2029

\bibitem[\protect\citeauthoryear{{Townsend} \& {Teitler}}{{Townsend} \&
  {Teitler}}{2013}]{gyre1}
{Townsend} R.~H.~D.,  {Teitler} S.~A.,  2013, \mn@doi [\mnras]
  {10.1093/mnras/stt1533}, \href
  {http://ukads.nottingham.ac.uk/abs/2013MNRAS.435.3406T} {435, 3406}

\bibitem[\protect\citeauthoryear{{Townsend}, {Goldstein}  \&
  {Zweibel}}{{Townsend} et~al.}{2018}]{gyre2}
{Townsend} R.~H.~D.,  {Goldstein} J.,   {Zweibel} E.~G.,  2018, \mn@doi
  [\mnras] {10.1093/mnras/stx3142}, \href
  {http://ukads.nottingham.ac.uk/abs/2018MNRAS.475..879T} {475, 879}

\bibitem[\protect\citeauthoryear{{Trampedach}, {Stein},
  {Christensen-Dalsgaard}, {Nordlund}  \& {Asplund}}{{Trampedach}
  et~al.}{2014a}]{trampedach2014a}
{Trampedach} R.,  {Stein} R.~F.,  {Christensen-Dalsgaard} J.,  {Nordlund}
  {\AA}.,   {Asplund} M.,  2014a, \mn@doi [\mnras] {10.1093/mnras/stu889},
  \href {http://esoads.eso.org/abs/2014MNRAS.442..805T} {442, 805}

\bibitem[\protect\citeauthoryear{{Trampedach}, {Stein},
  {Christensen-Dalsgaard}, {Nordlund}  \& {Asplund}}{{Trampedach}
  et~al.}{2014b}]{trampedach2014b}
{Trampedach} R.,  {Stein} R.~F.,  {Christensen-Dalsgaard} J.,  {Nordlund}
  {\AA}.,   {Asplund} M.,  2014b, \mn@doi [\mnras] {10.1093/mnras/stu2084},
  \href {http://esoads.eso.org/abs/2014MNRAS.445.4366T} {445, 4366}

\bibitem[\protect\citeauthoryear{{Valenti} \& {Fischer}}{{Valenti} \&
  {Fischer}}{2005}]{valenti2005}
{Valenti} J.~A.,  {Fischer} D.~A.,  2005, \mn@doi [\apjs] {10.1086/430500},
  \href {https://ui.adsabs.harvard.edu/abs/2005ApJS..159..141V} {159, 141}

\bibitem[\protect\citeauthoryear{{Weiss} \& {Schlattl}}{{Weiss} \&
  {Schlattl}}{2008}]{weiss2008}
{Weiss} A.,  {Schlattl} H.,  2008, \mn@doi [\apss] {10.1007/s10509-007-9606-5},
  \href {https://ui.adsabs.harvard.edu/abs/2008Ap&SS.316...99W} {316, 99}

\bibitem[\protect\citeauthoryear{{Wright} et~al.,}{{Wright}
  et~al.}{2010}]{wise}
{Wright} E.~L.,  et~al., 2010, \mn@doi [\aj] {10.1088/0004-6256/140/6/1868},
  \href {https://ui.adsabs.harvard.edu/abs/2010AJ....140.1868W} {140, 1868}

\bibitem[\protect\citeauthoryear{{da Silva}, {Milone}  \& {Rocha-Pinto}}{{da
  Silva} et~al.}{2015}]{dasilva2015}
{da Silva} R.,  {Milone} A. d.~C.,   {Rocha-Pinto} H.~J.,  2015, \mn@doi [\aap]
  {10.1051/0004-6361/201525770}, \href
  {https://ui.adsabs.harvard.edu/abs/2015A&A...580A..24D} {580, A24}

\makeatother
\end{thebibliography}
\input{HR1988.bbl}

\bsp	
\label{lastpage}
\end{document}